\documentclass[fleqn,usenatbib]{mnras}

\usepackage{newtxtext,newtxmath}

\usepackage[T1]{fontenc}

\DeclareRobustCommand{\VAN}[3]{#2}
\let\VANthebibliography\thebibliography
\def\thebibliography{\DeclareRobustCommand{\VAN}[3]{##3}\VANthebibliography}

\usepackage{graphicx}	
\usepackage{amsmath}	
\usepackage{siunitx}
\usepackage{multirow}
\usepackage{gensymb}
\usepackage{float}
\usepackage[normalem]{ulem}
\usepackage{mathtools}
\usepackage{orcidlink}
\usepackage{float}
\usepackage[flushleft]{threeparttable}

\defcitealias{KLI}{KL-I}

\title[Cluster Lensing with $\mathcal{O}(1)$ Galaxies]{Kinematic Lensing Inference II: Cluster Lensing with $\mathcal{O}(1)$ Galaxies}

\author[R. S. et al.]{Pranjal R. S.\orcidlink{0000-0003-3714-2574},$^{1}$\thanks{E-mail: pranjalrs@arizona.edu}
Eric Huff\orcidlink{0000-0002-9378-3424},$^2$
Elisabeth Krause\orcidlink{0000-0001-8356-2014},$^{1,3}$
Tim Eifler\orcidlink{0000-0002-1894-3301},$^{1,3}$
Spencer Everett\orcidlink{0000-0002-3745-2882},$^4$
\newauthor
Yu-Hsiu Huang\orcidlink{0000-0002-4982-0208},$^1$
Jiachuan Xu\orcidlink{0000-0003-0871-8941}$^1$
\\
$^{1}$Department of Astronomy/Steward Observatory, University of Arizona, 933 North Cherry Avenue, Tucson, AZ 85721, USA\\
$^{2}$Jet Propulsion Laboratory, California Institute of Technology, Pasadena, CA 91109, USA\\
$^3$Department of Physics, University of Arizona, 1118 E Fourth Street, Tucson, AZ 85721, USA\\
$^4$California Institute of Technology, 1200 E California Blvd, 91125, USA
}

\date{Accepted XXX. Received YYY; in original form ZZZ}

\pubyear{2015}

\begin{document}
\label{firstpage}
\pagerange{\pageref{firstpage}--\pageref{lastpage}}
\maketitle

\begin{abstract}
We present the first detection of a cluster lensing signal with `Kinematic Lensing' (KL), a novel weak lensing method that combines photometry, spectroscopy, and the Tully-Fisher relation to enable shear measurements with individual source galaxies. This is the second paper in a two-part series aimed at measuring a KL signal from data. The first paper, \citet{KLI}, describes the  inference pipeline, which jointly forward models galaxy imaging and spectroscopy, and demonstrates unbiased shear inference with simulated data. This paper presents measurements of the lensing signal from the galaxy cluster Abell 2261. We obtain spectroscopic observations of background disk galaxies in the cluster field selected from the CLASH Subaru catalog. The final sample consists of three source galaxies while the remaining are rejected due to insufficient signal-to-noise, spectroscopic failures, and inadequately sampled rotation curves. We apply the KL inference pipeline to the three sources and find the shear estimates to be in broad agreement with traditional weak lensing measurements. The typical shear measurement uncertainty for our sources is $\sigma(g_+)\approx 0.026$, which represents approximately a ten-fold improvement over the weak lensing shape noise. We identify target selection and observing strategy as the key avenues of improvement for future KL programs.
\end{abstract}




\section{Introduction}
Weak lensing (WL) --- the deflection of photon trajectories due to tidal fields --- serves as a key observational probe of the distribution of luminous and dark matter.
This effect introduces coherent distortions in observed shapes of background source galaxies, which have been utilized by photometric surveys to place tight constraints on cosmological parameters \citep[e.g.][]{Asgari2021,Amon2022,Li2023, Abott2023}. Weak lensing is the primary science driver for Stage IV programs such as \textit{Euclid}\footnote{\href{https://sci.esa.int/web/euclid}
{\nolinkurl{https://sci.esa.int/web/euclid}}} \citep{Euclid}, the Rubin Observatory’s Legacy Survey of Space and Time (LSST\footnote{\href{https://www.lsst.org}{\nolinkurl{https://www.lsst.org}}}, \citealt{LSST}), and the \textit{Nancy Grace Roman Space Telescope} (\textit{Roman}\footnote{\href{https://roman.gsfc.nasa.gov}{\nolinkurl{https://roman.gsfc.nasa.gov}}}, \citealt{Roman}). These surveys are set to boost measurement precision and significantly improve cosmological constraints from weak lensing.

Weak lensing in all its forms such as cosmic shear, galaxy-galaxy lensing, and cluster-galaxy lensing, is fundamentally a statistical measurement. The main source of statistical uncertainty stems from the unknown intrinsic galaxy shape, also known as shape noise ($\sigma_\epsilon\approx0.3$ per component), which dominates over the gravitational shear signal. Thus for an individual source galaxy, it is impossible to differentiate between the effects of weak lensing and the intrinsic shape using only photometry. As a consequence of this shape-shear degeneracy, traditional weak lensing requires large galaxy samples in order to extract the shear signal from an inherently noisy measurement. Moreover, the inclusion of faint galaxies in these samples makes it challenging to model systematics arising from photometric redshift and shape measurement uncertainties.

Several studies have discussed the use of disk galaxy kinematics to disentangle the intrinsic shape from the WL signal \citep{Blain2002,Morales2006,2015MNRAS.451.2161D,DiGiorgio2021}. Rather than relying on distortions in galaxy shapes, these techniques utilize the breaking of symmetry in the velocity field to achieve high-precision shear measurements. \citet{Gurri2020} conducted the first such measurement by analyzing integral field unit (IFU) observations of 18 sources in a galaxy-galaxy lensing geometry. They achieve a mean shear measurement of $0.020\pm0.008$, which demonstrates the potential of this approach to overcome the limitations faced by traditional weak lensing.

\citet{Huff2013} proposed a technique, known as kinematic lensing (KL), which uses spectroscopic observations and the Tully-Fisher (TF) relation \citep{Tully1977} to reduce shape noise. KL infers the intrinsic galaxy shape, which in combination with the observed shape, breaks the shape-shear degeneracy. Our team's forecasts suggest that KL can achieve shape noise in the range $\sigma_\epsilon=0.02-0.04$ \citep{KLI,Xu2023}.
This reduction in shape noise enhances the signal-to-noise (S/N) per source to $\sim1$, which means that KL can achieve the same precision as traditional WL but with a $100\times$ smaller sample. The increased statistical power also enables us to preferentially target bright and well-resolved source galaxies that provide better shape measurements. Moreover, incorporating spectroscopic data in the inference eliminates  photometric redshift uncertainties, a major challenge for WL surveys \citep{Schmidt2020, Zhang2023}. 
KL is also robust to intrinsic alignments (IA) and related astrophysical analogs \citep{Huang2024}. The increased shear-measurement signal-to-noise per source galaxy and robustness to key WL systematics factors suggest that KL can be a promising probe for wide-field spectroscopic surveys \citep{Xu2023, Xu2024}.

This paper is the second in a two-part series aimed at  a proof of concept measurement with KL. In the first paper (\citealt{KLI}, hereafter KL-I) we describe the implementation of the inference pipeline, demonstrate unbiased shear inference on simulated data at realistic signal-to-noise, and characterize the dependence of shape noise on observational factors. This second paper extends our previous work for application to real data and presents a pilot cluster lensing measurement. Cluster lensing \citep[see][for review]{Umetsu2020} provides a strong shear signal which has been well characterized by weak lensing analyses \citep{WtG2014a,Okabe2016, Umetsu2016,Klein2019,Murray2022,Grandis2024}. Specifically, our target cluster Abell 2261 has been extensively studied under the Cluster Lensing and Supernova survey with Hubble (CLASH, \citealt{Postman2012, Umetsu2014}) program, which provides a point of comparison for our measurement.

The paper is organized as follows. In Sec. \ref{sec:kl_theory}, we review the kinematic lensing method. Section \ref{sec:deimos_data} details the data products used in this work, including the identification of targets from photometric catalogs and the strategy for spectroscopic follow-up. Section \ref{sec:pipeline} outlines the shear inference methodology, including the forward model for image and spectrum, stellar mass estimation, and likelihood analysis. We present results of our KL inference and comparison with existing measurements in Sec. \ref{sec:measurement} . In Sec. \ref{sec:conclusions}, we present our conclusions and outlook for future KL observations.

\section{Theoretical Background}
\label{sec:kl_theory}
\begin{figure}
    \centering
    \includegraphics[width=\linewidth]{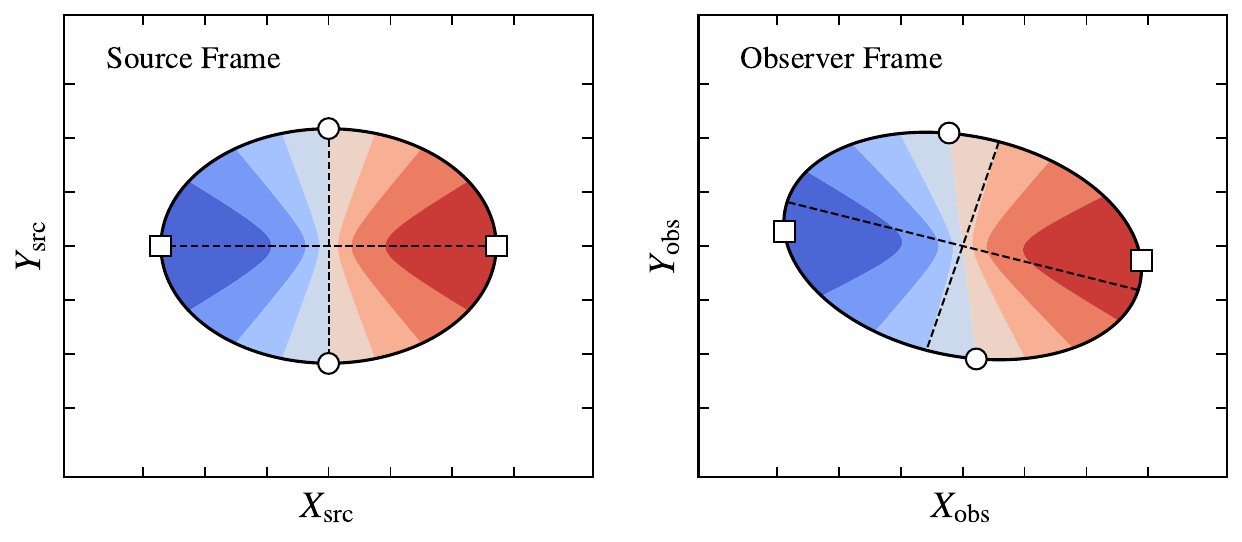}
    \caption{Illustration of the effect of weak lensing on galaxy photometry and kinematics. Left: In the source frame, the velocity field (filled contours) is symmetric about the photometric major and minor axis (dashed black lines). Therefore, the major and minor axes also correspond to the locations of the maximum (open squares) and zero (open circles) line-of-sight velocity, respectively. Right: The application of shear distorts the isophote and breaks the symmetry in the velocity field. As a result, we measure a nonzero line-of-sight velocity along the photometric minor axis while the velocity along the major axis is less than the maximum line-of-sight velocity.}
    \label{fig:KL_illustration}
\end{figure}
Gravitational lensing occurs when the inhomogeneous distribution of matter along the line of sight influences the propagation of photons from background sources.
This effect results in the deflection of photon trajectories while their specific intensity and wavelength are preserved. In the WL regime, the distortion in observed shapes occur because the deflection varies across galaxy image; and the gradients in deflections are governed by the lensing distortion matrix \textsf{\textbf{A}}. Thus, the relationship between a point in the source frame $\vec{\zeta}_\mathrm{src}$ and its observed position $\vec{\zeta}_\mathrm{obs}$ is expressed as
\begin{equation}
    \vec{\zeta}_\mathrm{obs} = \textsf{\textbf{A}}^{-1} \vec{\zeta}_\mathrm{src},
\end{equation}
\begin{equation}
\label{eq:A_matrix}
\textsf{\textbf{A}} =
(1-\kappa)\begin{pmatrix}
1-g_1 & -g_2 \\
-g_2 & 1 + g_2
\end{pmatrix},
\end{equation}
where we define the observer and source frames as the reference frames at the positions of the observer and the source galaxy, respectively. The convergence $\kappa$ quantifies the isotropic change in size whereas the components of the reduced complex shear\footnote{The reduced shear is related to the shear, $\gamma$, as $g=\gamma/(1-\kappa)$. Assuming $\kappa\ll1$ implies $g\approx\gamma$, which is also referred to as the reduced shear approximation. Hereon, we use the term shear to refer to reduced shear unless stated otherwise.} $g = g_1 + ig_2$ represent the  distortion along the $x$-axis and the axis rotated by $45\degree$ from the $x$-axis, respectively.
 Since gravitational lensing preserves specific intensity, the observed intensity $I_\mathrm{obs}$ at position $\vec{\zeta}_\mathrm{obs}$ is related to the intensity  $ I_\mathrm{src}$ in the source frame as
\begin{equation}
\label{eq:obs_intensity}
    I_\mathrm{obs}(\vec{\zeta}_\mathrm{obs}) = I_\mathrm{src}(\textsf{\textbf{A}}\, \vec{\zeta}_\mathrm{obs}).
\end{equation}
The impact of this transformation on galaxy photometry is a stretching/squeezing along the major axis and an apparent rotation\footnote{\label{fn:gal_orientation}For Fig. \ref{fig:KL_illustration} and the remainder of this section, we assume that the major and minor axes of the galaxy in the source frame are aligned with the $x$ and $y$ axes, respectively. Therefore, the $g_1$ component is along the major axis and $g_2$ component is aligned $45\degree$ from the major axis.}, as illustrated in Fig. \ref{fig:KL_illustration}. The above relation implies that the observed galaxy ellipticity ($\epsilon_\mathrm{obs}$) is a function of \textit{intrinsic} ellipticity ($\epsilon_\mathrm{int}$) and gravitational shear, the two being completely degenerate. This relationship can be written as
\begin{equation}
    \epsilon_\mathrm{obs} = \epsilon_\mathrm{int} + g,
\end{equation}
where we express complex ellipticity as $\epsilon=\epsilon_1 + i \epsilon_2$, with the modulus equal to the scalar ellipticity $e=|\epsilon|$.
This fundamental limitation prevents traditional weak lensing, which relies solely on photometry, from measuring shear for individual galaxies. Instead, one has to average over observed shapes of a sufficiently large sample so that the intrinsic shape contributions cancel out (i.e. $\langle\epsilon_\mathrm{int}\rangle=0$), leaving only the coherent shear signal.

To enable inference on a per-galaxy basis, kinematic lensing leverages the fact that weak lensing has distinct effects on galaxy shapes and velocity fields. We can relate the observed velocity field $V_\mathrm{obs}$ to the line-of-sight velocity field in the source frame $V_\mathrm{src}$ as
\begin{equation}
    V_\mathrm{obs}(\vec{\zeta}_\mathrm{obs}) = V_\mathrm{src}(\textsf{\textbf{A}}\, \vec{\zeta}_\mathrm{obs}).
\end{equation}
As shown in the left panel of Fig. \ref{fig:KL_illustration}, the velocity field of an unsheared source is symmetric about the photometric major and minor axis. This symmetry is broken when shear, more specifically the $g_2$ component, is applied (as stated in footnote \ref{fn:gal_orientation} the $g_2$ component is aligned $45\degree$ from the major axis). As a result, the photometric minor axis in the observer frame does not correspond to a zero line-of-sight velocity. Similarly, the location of the maximum line-of-sight velocity is not mapped onto the observed photometric major axis.

The deviation of disk kinematics from symmetry is a clear signature of the $g_2$ shear component and was first discussed by \citet{Blain2002} and \citet{Morales2006} in the context of radio telescopes. The formalism was extended to IFU observations in \citet{2015MNRAS.451.2161D} and has been used for galaxy-galaxy lensing measurements in \citet{Gurri2020}. As opposed to IFU data, single-slit observations like the ones used in this work do not contain sufficient information to discern symmetry (or lack thereof) in the velocity field. In such a scenario, it is the kinematic misalignment i.e. misalignment between the kinematic and photometric axes that provides constraining power.

On the other hand, the $g_1$ component --- which causes a stretching or squeezing of the velocity field along the major axis --- cannot be estimated through its effect on the kinematics and we require additional information in the form of the Tully-Fisher relation  \citep{Tully1977}. The TF relation is empirically determined by measuring rotation curves of a galaxy population and predicts the maximum circular velocity $V_\mathrm{TF}$ as a function of stellar mass $M_\star$
\begin{equation}
\label{eq:TF_relation}
    \log V_\mathrm{TF} = a + b \log M_\star,
\end{equation}
where $a$ and $b$ are the intercept and slope of the relation, respectively. To understand the role of the TF relation, consider a source galaxy for which we measure the line-of-sight velocity along the photometric major and minor axes as $V_\mathrm{major}$ and $V_\mathrm{minor}$, respectively. Since $V_\mathrm{major}$ is simply a line-of-sight projection of $V_\mathrm{TF}$, we can estimate the galaxy inclination $i$ and intrinsic scalar ellipticity $e_\mathrm{int}$ as
\begin{equation}
    \sin \hat{i} = \frac{V_\mathrm{major}}{V_\mathrm{TF}}\,; \quad
    e_\mathrm{int} = \frac{1-\sqrt{1 - (1-q_\text{z}^2) \sin^2\hat{i}}}{1+\sqrt{1 - (1-q_\text{z}^2) \sin^2\hat{i}}},
\end{equation}
where  $q_\text{z}$ is the edge-on aspect ratio. Thus, the deviation of a galaxy from the population average i.e. the TF relation enables us to link the observed kinematics to its intrinsic galaxy shape. We can estimate the shear components $g_{1,2}$ by combining $e_\mathrm{int}$ with the observed ellipticity $e_\mathrm{obs}$ \citep[see][for details]{Xu2023}
\begin{align}
\label{eq:g1_estimator}
    \hat{g}_1 &= \frac{e_\mathrm{obs}^2-e_\mathrm{int}^2}{2e_\mathrm{obs}^2(1-e_\mathrm{int}^2)}, \\\label{eq:g2_estimator}
    \hat{g}_2 &= \left|\frac{V_\mathrm{minor}}{V_\mathrm{major}}\right|\frac{2e_\mathrm{int}}{\cos\hat{i}(1+2e_\mathrm{int}+e_\mathrm{obs}^2)}.
\end{align}
The above expression demonstrates how the $g_1$ component can be estimated from knowledge of the intrinsic galaxy shape. It also shows how the second shear component can be inferred from kinematic misalignment, as $\hat{g}_2\neq0$ when $V_\mathrm{minor}\neq0$. Note that we do not use the shear estimators in Eqs. (\ref{eq:g1_estimator}) \& (\ref{eq:g2_estimator}) for inference and instead take the forward modeling approach presented in Sec. \ref{sec:pipeline}.

In summary, the advantage of using kinematics is that disk galaxies (i) follow the well-known TF relation and (ii) their kinematics exhibit symmetry in the absence of shear. By combining these two characteristics, we can constrain the kinematics and shape of the unsheared source galaxy. Thus, the difference between the observed and expected kinematics and shape enables us to estimate the weak lensing shear with high precision.

\section{Spectroscopic Data}
\label{sec:deimos_data}
\begin{figure}
    \centering
\includegraphics[width=\linewidth]{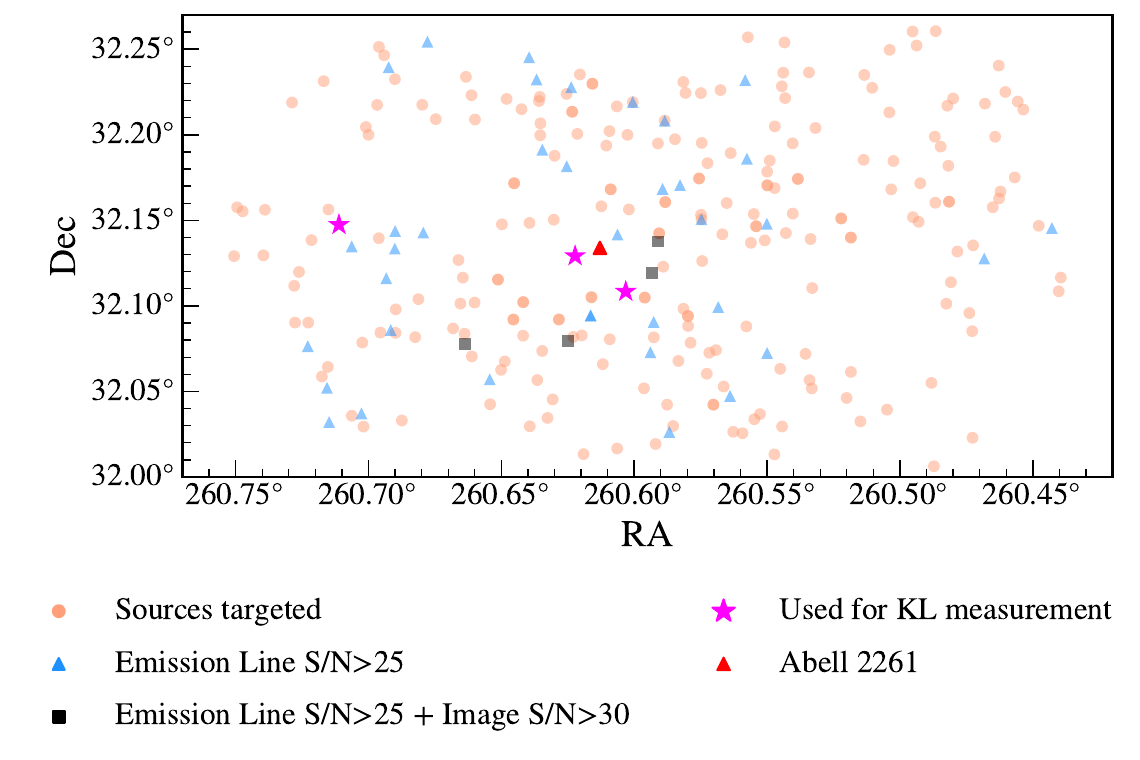}
    \caption{Distribution of background source galaxies around Abell 2261. We obtain DEIMOS observations for a total of 141 sources out which three (shown as star symbols) are used for the KL measurement. We refer the reader to Sec. \ref{sec:sample_selection} for details about down-selection from the targeted sources.}
    \label{fig:deimos_targets}
\end{figure}

\subsection{KL source selection}
We selected bright, well-resolved disk galaxies at redshifts well behind our target cluster, Abell 2261 (abbreviated as A2261) at $z=0.224$.
The CLASH program \citep{Umetsu2014,2017MNRAS.470...95M} provides photometric redshifts and spectral template fits using the Bayesian Photo-z (BPZ) software \citep{BPZ} and Subaru photometry. However, we needed to derive additional morphological information from the imaging to effectively target disk galaxies.
To do this, we selected sources from the available CLASH data products with redshifts above $z>0.3$, and best-fit spectral templates corresponding to star-forming galaxies. We derived galaxy morphological measurements suitable for these potential targets from the available CLASH Subaru $i$-band imaging for the cluster. We used PSFEx \citep{Bertin2011} to build a spatially-varying PSF model, and then used SEXtractor \citep{Bertin1996} to fit PSF-convolved S\'ersic models to each source that passed the photometric selection based on CLASH photometry. We then excluded any galaxies with a  S\'ersic index $n>2$ or a half-light radius $r<1.5''$, as well as those where visual inspection indicated that a target's photometry was significantly impacted by blending with neighbors. The remaining galaxies we considered eligible spectroscopic targets.

\subsection{Observing strategy and data reduction}
The spectroscopic observations used in this work are from  DEep Imaging Multi-Object Spectrograph (DEIMOS, \citealt{Faber2003}), a multi-object spectrograph on the Keck-II telescope. The observations were taken May-June 2014. To resolve the [OII] doublet and obtain accurate kinematics from $0.6<z<1.2$, we used the 1200 line/mm grating centered $ \lambda = 7200$ {\AA}.  This yielded a spectral resolution of $1.7$ \AA, or $30$ km/s. As our targets were well-resolved disks (half-light radii $>1.5''$), we selected a $1''$ slit width in order to maximize the signal while still permitting accurate capture of the resolved disk kinematics.

The DEIMOS slitmask sky footprint is $4'\times 16'$. We observed the cluster through four separate slitmasks, each half-covering the cluster, and oriented at $0$, $90$, $180$, and $270$ degrees east of north in order to ensure even coverage of an approximately $8'\times 8'$ square region centered on A2261, shown in Fig. \ref{fig:deimos_targets}. After excluding alignment stars, we obtained spectra for 141 source galaxies. The 2D spectra are reduced using the \textsc{spec2d} pipeline used in the DEEP2 survey \citep{deep2}. We extract the one-dimensional spectrum using the \textsc{spec1d} package \citep{2003SPIE.4834..161D} and estimate redshifts using the \textsc{zspec} fitting package, which interactively fits a redshift and a combination of galaxy and stellar spectra templates \citep{2007ApJ...665..265F,2004ApJ...609..525C}. For each galaxy spectrum, we generate 3 {\AA} cutouts centered on the emission lines of interest. For each cutout we subtract the continuum emission which we estimate by fitting a Gaussian as a function of spatial position.

\begin{table*}
\begin{tabular}{ l l l l}
\hline\hline
Parameter & Description & Prior & Units\\
\hline
\multicolumn{4}{c}{ \textbf{Shared parameters}} \\
$g_+$ & Tangential shear & Uniform $(-0.5, 0.5)$ & -\\
$\cos i$ & Inclination &  Uniform (0, 1) & -\\
$\theta_\mathrm{int}$ & Intrinsic galaxy position angle & Uniform $(0,2\pi)$ & radian \\
\hline \multicolumn{4}{c}{ \textbf{Spectrum model}} \\
$V_\mathrm{circ}$ & Maximum circular velocity & Gaussian$(\log V_\mathrm{TF}, \sigma_\mathrm{TF})$ & km s$^{-1}$\\
$r_\mathrm{vscale}$ & Velocity scale radius &  Uniform (0.1, 3)& arcsec\\
$V_\mathrm{sys}$ & Systemic velocity & Uniform$(-500, 500)$ & arcsec\\
$\Delta x_\mathrm{kin}$ & Kinematic offset & Uniform (-0.3, 0.3) & Fraction of disk size\\
$\Delta y_\mathrm{kin}$ & Kinematic offset & Fixed to 0 & Fraction of disk size\\
$I_\mathrm{spec}$ & Spectrum central brightness & Uniform $(0, 500)$ & Arbitrary units\\
$B_\mathrm{spec}$ & Spectrum background level & Uniform $(-10, -10)$ & Arbitrary units\\
\hline \multicolumn{4}{c}{ \textbf{Image: bulge $+$ disk model}} \\
$n_\mathrm{Sersic}$ & Disk S\'ersic index & Fixed to 1 & -\\
$r_\mathrm{hl,disk}$ & Disk half-light radius & Uniform (0.15, 3)& arcsec\\
$\Delta x_{\mathrm{disk}}$ & Disk offset& Uniform (-0.3, 0.3) & Fraction of disk size\\
$\Delta y_\mathrm{disk}$ & Disk offset& Uniform (-0.3, 0.3) & Fraction of disk size\\
$F_\mathrm{disk}$ & Total disk flux & Log Uniform ($1,10^5$)& Arbitrary units\\
$q_\text{z}$ & Disk edge-on aspect ratio & Fixed to 0.2 & - \\
$r_\mathrm{hl,bulge}$  &Bulge half-light radius & Uniform (0.15, 3)& arcsec\\
$\Delta x_\mathrm{bulge}$ & Bulge offset& Uniform (-0.3, 0.3)& Fraction of disk size \\
$\Delta y_\mathrm{bulge}$ & Bulge offset& Uniform (-0.3, 0.3)& Fraction of disk size \\
$F_\mathrm{bulge}$ & Total bulge flux &  Log Uniform ($1,10^5$)& Arbitrary units\\
\hline \multicolumn{4}{c}{ \textbf{Image: disk only model}} \\
$n_\mathrm{Sersic}$ & Disk S\'ersic index & Uniform $(0.3, 1.5)$ & -\\
$r_\mathrm{hl}^{\mathrm{disk}}$ & Disk half-light radius & Uniform (0.15, 3)& arcsec\\
$\Delta x_{\mathrm{disk}}$ & Disk offset& Uniform (-0.3, 0.3) & Fraction of disk size\\
$\Delta y_\mathrm{disk}$ & Disk offset& Uniform (-0.3, 0.3) & Fraction of disk size\\
$F_\mathrm{disk}$ & Total disk flux & Log Uniform ($1,10^5$)& Arbitrary units\\
$q_\text{z}$ & Disk edge-on aspect ratio & Fixed to 0.2 & - \\
\hline\hline
\end{tabular}

\caption{Parameters and priors for the joint image and spectrum model. The shared parameters apply to both the spectrum and image models, while the remaining parameters are specific to each model.}
\label{tab:model_params}
\end{table*}

\section{KL Inference Framework}
\label{sec:pipeline}
As described in Sec. \ref{sec:kl_theory}, the combination of imaging and spectroscopy breaks the shape-shear degeneracy and enables shear inference with individual sources. For KL inference, we build a joint forward model of the source galaxy image and spectrum. This model includes shared parameters for shear and galaxy properties, along with parameters specific to the image and spectrum as detailed in Table \ref{tab:model_params}. The forward model is largely based on \citetalias{KLI} with the following changes:
\begin{itemize}
\item For the image model, we replace a simple exponential brightness profile for the disk with a bulge+disk decomposition or a S\'ersic profile with varying index, and incorporating additional parameters to control their 2D spatial offsets.
\item In the spectrum model, rather than relying on simplified assumptions of an exponential brightness profile and a constant emission line width, we estimate these directly from the data. The model also includes parameters governing the offset between the kinematic and photometric centers.
\end{itemize}

For completeness we summarize the updated image and spectrum models below.

\subsection{Image model}
We forward model the image using the open-source software \textsc{Galsim} \citep{RJM+15}. We use two methods for modeling the intensity profile (a) decomposing it as a sum of a disk component with an exponential profile and a bulge component which follows a de Vaucouleurs profile or (b) representing the galaxy as a disk with a general S\'ersic profile. For each source galaxy we choose the method which provides a better goodness of fit.

In both approaches, the 3D intensity profile of the disk is characterized by: the S\'ersic index $n$ ($n=1$ for an exponential profile), the half-light radius $r_\mathrm{hl,disk}$, the edge-on aspect ratio $q_\text{z}$ and the total flux $F_\mathrm{disk}$. We project the intensity profile onto to the 2D plane based on the the inclination $i$, apply a shift by ($\Delta x_\mathrm{disk}$, $\Delta y_\mathrm{disk}$), and a rotation by the intrinsic position angle $\theta_\mathrm{int}$. The 2D intensity profile of the bulge component is specified by its size $r_\mathrm{hl,bulge}$, total flux $F_\mathrm{bulge}$ and offset parameters ($\Delta x_\mathrm{bulge}$, $\Delta y_\mathrm{bulge}$). We truncate the intensity profile for both components at four times their respective half light radii. We shear the total intensity distribution by  $(g_1,g_2)$ and convolve with a Gaussian PSF with a full-width half maximum (FWHM) equal to the observation seeing. The final image is rendered on a grid with the pixel size set by the instrument resolution.

\subsection{Spectrum model}
\subsubsection{Kinematics Model}
\label{sec:kin_model}
We assume that the kinematics of disk galaxies can be approximated by a circularly rotating, infinitesimally thin disk. There are several parametric forms used to describe rotation curves \citep[e.g. ][]{Persic1996, Courteau1997}, here we choose the commonly used $\arctan$ function. In the reference frame where the galaxy is a face-on disk, the circular velocity at point with polar coordinates $(r,\phi)$ is given by
\begin{equation}
\label{eq:vel_profile}
    V(r,\phi) = \frac{2}{\pi} V_\mathrm{circ} \cos (\phi) \arctan\left(\frac{r}{r_\mathrm{vscale}}\right).
\end{equation}
Here, $V_\mathrm{circ}$ is the maximum circular velocity and $r_\mathrm{vscale}$ is the velocity scale radius (also refered to as the turnover radius) which marks the transition between the rising and flattening part of the rotation curve. The prior on $V_\mathrm{circ}$ is informed by the TF relation and the galaxy stellar mass (refer to Secs. \ref{sec:Mstar_est} and \ref{sec:param_inference}). 

At a point $(r',\phi')$ in the observer frame, the line-of-sight projected velocity is
\begin{equation}
    \label{eq:v_obs}
    V_\mathrm{obs}(r',\phi') = V_\mathrm{sys} + V(r,\phi)\times\sin i,
\end{equation}
where $V_\mathrm{sys}$ is the  galaxy systemic velocity and $i$ is the galaxy inclination; $i=0$ corresponds to a face-on disk while $i=\pi/2$ is an edge-on disk. As the kinematic and photometric center can be offset from each other, we also apply a shift ($\Delta x_\mathrm{kin},\Delta y_\mathrm{kin}$) to the velocity field $V(r,\phi)$.

The relation between the observed coordinates $\vec{\zeta}'=r'(\cos\phi',\sin\phi')^{\text{T}}$ and the face-on coordinates $\vec{\zeta}=r(\cos\phi,\sin\phi)^{\text{T}}$ is expressed as
\begin{equation}
\label{eq:coord_mapping}
\vec{\zeta}'= \textsf{\textbf{A}}^{-1}\text{ }\textsf{\textbf{R}}\text{ }\textsf{\textbf{I}}\text{ }\vec{\zeta}.
\end{equation}
The matrices in the above equation represent a series of reference frame transformations: \textsf{\textbf{I}} represents the projection due to the intrinsic galaxy inclination, \textsf{\textbf{R}} represents the rotation to account for the intrinsic galaxy position angle and the lensing distortion matrix \textsf{\textbf{A}} accounts for the effect of shear. The lensing distortion matrix is defined in Eq. (\ref{eq:A_matrix}), the remaining two matrices are defined as

\begin{align}
\textsf{\textbf{R}} = 
\begin{pmatrix} 
\cos \theta & -\sin \theta \\
\sin \theta & \cos \theta
\end{pmatrix}; \hspace{3pt}
\textsf{\textbf{I}} &= 
\begin{pmatrix} 
1 & 0 \\
0 & \cos i
\end{pmatrix},
\end{align}
where $\theta$ is the rotation angle.

\subsubsection{Emission line model}
To convert galaxy kinematics into an emission line feature, we require: emission line wavelength in the observer frame, line width, and intensity. We compute the observed emission line wavelength from the line-of-sight velocity field computed in Eq. (\ref{eq:v_obs})
\begin{equation}
\lambda_\text{obs}\left(r',\phi'\right)=(1+z)\left(1+\frac{V_\mathrm{obs}\left(r',\phi'\right)}{c}\right) \lambda_0,
\end{equation}
where  $\lambda_0$ is the rest-frame emission wavelength, $z$ is the galaxy redshift and $c$ is the speed of light.

\citetalias{KLI} assumed that the instrument's spectral resolution dominates the observed line width, which may not be valid in practice due to thermal/pressure broadening and environment-dependence of these line broadening mechanisms.  Additionally, the simplified assumption of an exponential brightness profile may not reflect the morphological substructure of disk galaxies.

To address these issues, we extract the emission line properties from the data and then use them in the model during the inference step. We smooth the \textit{observed} 2D spectrum by applying a median filter and fit a position dependent line profile. The line profile is composed of two half-Gaussians joined at the same peak i.e. the two halves have the same mean and amplitude but can have different widths. For each position along the slit, we fit this line profile using a least squares optimizer with the mean, amplitude, and widths as free parameters.

The position dependent amplitude and width are used to assign an intensity and line width to each pixel in the galaxy. The intensity of each pixel is scaled by $I_\text{spec}$, which determines the central brightness. We include
a flat background level $B_\text{spec}$ to account for any residual noise after sky subtraction. The intensity profile estimated from the observed 2D spectrum is already smeared by the PSF and we do not perform an additional PSF convolution on the data cube\footnote{In reality, the PSF smearing occurs before the spectral dispersion but this is likely a small effect.}.

The resulting 3D data cube has two axes corresponding to the spatial position on the sky and a third axis corresponding to the observed wavelength. Finally, we impose a slit mask on the 3D data cube and integrate across the shorter edge of the slit to obtain the 2D slit spectrum.

\subsection{Stellar mass estimation}
\label{sec:Mstar_est}
As described in Sec. \ref{sec:kl_theory}, we can disentangle the shape and kinematic information by placing a prior on the maximum circular velocity using the TF relation (Eq. \ref{eq:TF_relation}) and the galaxy stellar mass $M_\star$.

To estimate the galaxy stellar mass we use photometry from the Tractor catalog \citep{Lang2016} in DR10 of the Dark Energy Spectroscopic Instrument (DESI) Legacy Image Surveys \citep{Dey2019}. The catalog contains de-reddened fluxes in three DECam (Dark Energy Camera, \citealt{DECam}) filters $grz$ as well as the four WISE (Wide-field Infrared Survey Explorer, \citealt{WISE}) filters $W_1,W_2,W_3$ \& $W_4$.

We use the Python package $\textsc{Prospector}$ \citep{Johnson2021} to fit a galaxy spectral energy distribution (SED) to the observed photometry. We employ a delayed $\tau-$model for the star formation history (SFH) and a \citet{Chabrier2003} initial mass function (IMF). We model the galactic dust as a simple dust screen affecting stars of all ages equally with a \citet{Kriek2013} attenuation curve. Table \ref{tab:SED_params} lists the priors for the six varied parameters: \textit{surviving} stellar mass, metallicity, $e$-folding time for the SFH, age of stellar population, normalization of the dust attenuation curve, and power law of the dust attenuation curve. The remaining model parameters are fixed to their default values in the Flexible Stellar Population Synthesis for Python (FSPS\footnote{\href{https://dfm.io/python-fsps}
{\nolinkurl{https://dfm.io/python-fsps}}}, \citealt{Conroy2009, Conroy2010}) package.
\begin{table}
    \begin{tabular}{l l l}
    \hline\hline
        Parameter &  Description & Prior\\
         \hline
         $\log M_\star/\mathrm{M}_\odot$ & Surviving stellar Mass & Uniform (6,12)\\
         $\log Z/Z_\odot$ & Metallicity & Uniform (-2, 0.19)\\
         $\tau_\mathrm{SF}$ [Gyr] & SFH $e$-folding time & Log Uniform (0.1, 10)\\
         $t_\mathrm{age}$ [Gyr] & Age of stellar population& Uniform (0.001, 13.8)\\
         $\tau_{5500}$ & Dust attenuation norm. & Uniform (0, 4)\\
         $\nu$ & Dust attenuation slope & Uniform (-3, 1)\\
        \hline\hline
    \end{tabular}
    \caption{Model parameters and priors for fitting galaxy SED.}
    \label{tab:SED_params}
\end{table}

\subsection{Cluster lensing geometry}
\begin{figure}
    \centering
\includegraphics[width=\linewidth]{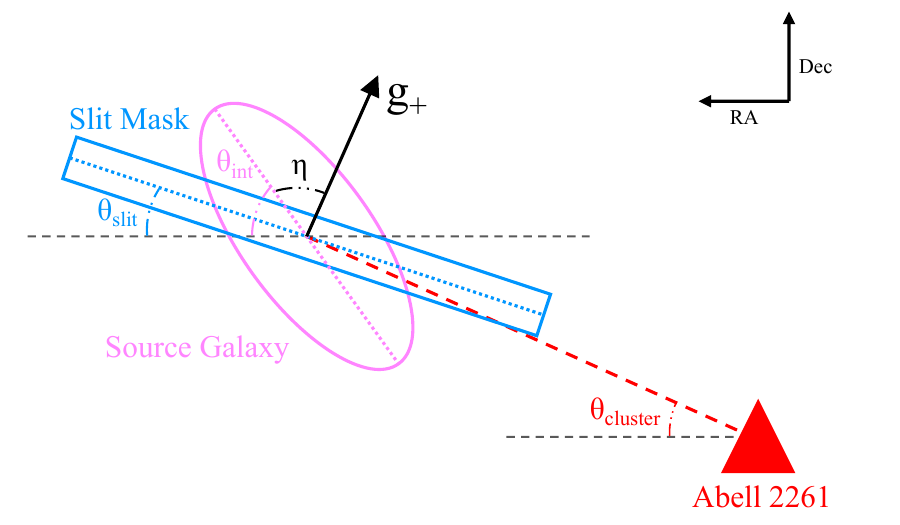}
    \caption{Illustration of the cluster lensing geometry.}
    \label{fig:lensing_geometry}
\end{figure}

The simulated KL inference demonstrated in \citetalias{KLI} utilizes two slit spectra, one aligned with the source galaxy major axis and the other aligned with the minor axis, to infer the two Cartesian shear components $g_{1,2}$. The Cartesian shears, at a location $\theta_\mathrm{cluster}$ around the cluster, are related to the tangential and cross components as
\begin{align}
    g_+ &= -(g_1\cos2\theta_\mathrm{cluster} + g_2 \sin2\theta_\mathrm{cluster}),\label{eq:gplus}\\
    g_\times &= -(g_2 \sin2\theta_\mathrm{cluster}-g_1\cos2\theta_\mathrm{cluster}).
\end{align}
However, the Abell 2261 observations consist of only one slit spectrum per galaxy, which only constrains a combination of the two shear components. Hence we assume an idealized cluster lensing geometry as illustrated in Fig. \ref{fig:lensing_geometry}, with all shear being tangentially aligned relative to the cluster center i.e. $g_\times=0$. This allows us to sample $g_+$ as a model parameter and evaluate  $g_{1,2}$ as
\begin{equation}
\label{eq:g_def}
    g_1 = -g_+\cos(2\theta_\mathrm{cluster})\,, \quad
    g_2 = -g_+\sin(2\theta_\mathrm{cluster})\,.
\end{equation}

The polar angle $\theta_\mathrm{cluster}$ is measured clockwise from the RA axis and is related to the lens coordinates $(\alpha_\text{l}, \delta_\text{l})$ and source coordinates $(\alpha_\text{s}, \delta_\text{s})$ as
\begin{equation}
    \theta_\mathrm{cluster} = \arctan\left[\frac{\delta_\text{s}-\delta_\text{l}}{(\alpha_\text{s}-\alpha_\text{l})\cos\left(\frac{\delta_\text{s}+\delta_\text{l}}{2}\right)}\right].
\end{equation}

\subsection{Parameter inference}
\label{sec:param_inference}
We derive posterior probability distributions using the \textsc{UltraNest}\footnote{\url{https://johannesbuchner.github.io/UltraNest/}} package \citep{buchner2021ultranest} that is based on the nested sampling Monte Carlo algorithm MLFriends \citep{Buchner2014, Buchner2019}. We assume a Gaussian likelihood for the joint image and spectrum model

\begin{align}
    \ln \mathcal{L} = - \frac{1}{2} \left[\sum_k\left(\frac{\mathrm{D}_k^{\mathrm{image}} - \mathrm{M}_k^{\mathrm{image}}}{\sigma_k^\mathrm{image}}\right)^2 + 
    \sum_k\left(\frac{\mathrm{D}_k^\mathrm{spec} - \mathrm{M}_k^\mathrm{spec}}{\sigma_k^\mathrm{spec}}\right)^2 \right],
\end{align}
where the index $k$ runs over all the pixels in the image 
or the spectrum.
In the above expression, $\mathrm{D}$, $\mathrm{M}$, and $\sigma$ represent the data, model, and noise, respectively. The superscripts indicate whether these quantities pertain to the image or the spectrum. The noise variance in the image (or spectrum) is the sum of the background sky variance and a Poisson noise term, $(\sigma_k^\mathrm{Poisson})^2 = \mathrm{D}_k^\mathrm{image}(\text{or}\,\mathrm{D}_k^\mathrm{spec})$. However, we find our observations to be background dominated and therefore the Poisson noise term has a small contribution.

We use the TF relation from \citet{Miller2011} for setting a Gaussian
prior on $\log V_\text{circ}$ centered at $\log V_\text{TF}$, where the latter is computed from Eq. (\ref{eq:TF_relation}) using intercept $a=1.718$ and slope $b=3.869$. The prior width ($\sigma_\text{TF}$) is the quadrature sum of the intrinsic dispersion in the TF relation $\sigma_\text{TF, int}=0.058$ dex and the uncertainty in $\log V_\text{TF}$ from stellar mass estimation. We use uniform priors for the remaining parameters. We restrict \{$\cos i, \theta_\mathrm{int}$\} to their mathematical range and set priors on \{$g_+,V_\mathrm{sys}$\} to ensure they have physically reasonable values. For $F_\mathrm{disk}$, we set a sufficiently wide prior based on the distribution of source fluxes from aperture photometry; we adopt the same prior for $F_\mathrm{bulge}$. Similarly, we set broad priors on \{$I_\text{spec}$, $B_\text{spec}$\} based on the distribution of maximum and minimum pixel values in the 2D spectra, respectively. The offset parameters are restricted to 30\% of the disk size as we do not expect larger shifts from the galaxy center. The upper bounds for the size parameters (e.g. $r_\mathrm{hl}^\mathrm{disk}$, $r_\mathrm{vscale}$) are based on examination of the source images which only span a few arcseconds, while the lower bounds are determined by the spatial resolution of the instrument. In preliminary analyses we found that it is not possible to constrain both kinematic offset parameters with a single slit measurement as the two are highly degenerate with each other and with the systemic velocity. Therefore, we only vary $\Delta x_\mathrm{kin}$ and fix $\Delta y_\mathrm{kin}$ to zero. For doublets like [OII] which are composed of two closely spaced emission lines, we use the same rotation curve for both emission line features but adopt separate systemic velocity and kinematic offsets.

To determine the appropriate image model i.e. disk with a S\'ersic profile where the S\'ersic index is allowed to vary or bulge$+$disk decomposition, we first fit just the $R_{c}$ band image (assuming zero tangential shear) and select the model that provides a better goodness of fit. For the bulge+disk model, the bulge size and flux are constrained to be smaller than the disk counterparts as we expect the disk component to dominate based on our selection criteria. For the source b008, we found that the default priors for the bulge offset were too restrictive for a good image fit and we therefore adopt wider prior range of $(-1,1)$ for $\Delta x_\mathrm{bulge}$ and $\Delta y_\mathrm{bulge}$.

\begin{table*}
\begin{threeparttable}
    \begin{tabular}{l l l l l l l l l l l }
    \hline\hline
    Source ID & Redshift & RA & Dec & $\log (M_\star/\text{M}_\odot)$ & $R$\tnote{a} & $\theta_\mathrm{cluster}$\tnote{b} &  $\beta_\mathrm{KL}$\tnote{c} & Emission Line S/N & Image S/N & Image Model\\
     \hline
     b007 & 0.623 & 260.6032  & 32.1084 & 10.12 $\pm$ 0.09 & 1.60$^\prime$  & -107.89$\degree$ & 0.604 & 43 &  97  & Bulge + Disk\\
     b008 & 0.585 & 260.7111  & 32.1474 & 10.77 $\pm$ 0.12 & 5.05$^\prime$  & 9.25$\degree$ & 0.582&  34 & 72 & Bulge + Disk\\
     c007 & 0.594 & 260.6222  &32.1291 & 10.29 $\pm$ 0.13  & 0.55$^\prime$ & -31.51$\degree$ & 0.587  & 55 &  77 & Disk only \\
    \hline\hline\\
    \end{tabular}
    \begin{tablenotes}
    \item[a] distance from cluster center
    \item[b] position angle from cluster center measured relative to the RA axis
    \item[c] lensing depth as defined in Eq. (\ref{eq:lensing_depth})
    \end{tablenotes}
\end{threeparttable}
    \caption{Ancillary data for the KL source sample.}
    \label{tab:KL_final_sample}
\end{table*}

\begin{figure}
    \centering
    \includegraphics[width=\linewidth]{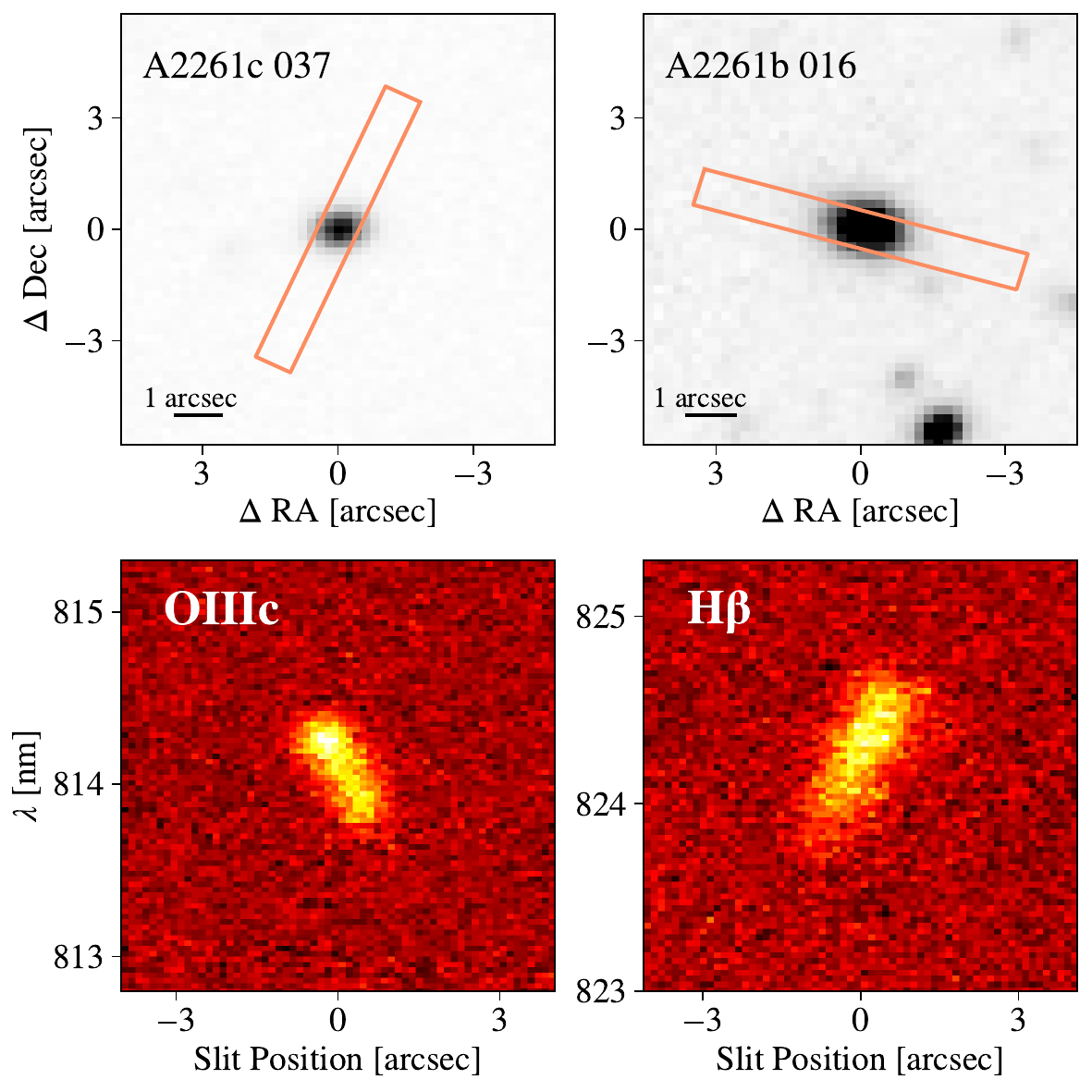}
    \caption{Examples of emission lines that do not meet the selection criterion. Top: Subaru $R_c$ band image and the slit orientation centered on the source. Bottom: The two emission lines are rejected as the characteristic `S' shape which corresponds to a flattened rotation curve is not detected. In each panel the $x$-axis is the position along the slit and $y$-axis is wavelength.}
    \label{fig:reject_example}
\end{figure}

\section{Measuring a KL signal - Cluster lensing from Abell 2261}
\label{sec:measurement}

\subsection{Sample selection}
\label{sec:sample_selection}
\begin{table}
    \begin{tabular}{l l c}
    \hline
    \hline
     & Criterion & \# of Sources \\
     \hline
       1. & Source galaxies targeted & 141 \\
       2. & Emission line S/N $>25$  & 44 \\
       3. & Excluding AGNs, sky line contamination, &  \multirow{2}{*}{24}\\
       & misidentified emission lines & \\
       4. & Image S/N$>30$ & 7\\
       5. & Visually examining rotation curve & 3\\
    \hline
    \hline
    \end{tabular}
    \caption{Number of source galaxies remaining at each step of the down-selection process.}
    \label{tab:selection}
\end{table}

Our prior forecasting work in \citetalias{KLI} suggests that accurate shear estimation requires high signal-to-noise emission lines (S/N$>25$), which originate in a disk component and are traceable out to the flattening of the disk rotation curve. Of the 141 target spectra, 44 have at least one emission line with S/N that meets this criterion. Excluding cases where emission originates from active galactic nuclei (AGN), where the emission line is misidentified or where there is a significant overlap with sky emission lines reduces this to 24 sources. We visually examine the remaining emission lines and reject cases in which the flattened rotation curve is not well measured in the disk outskirts. Figure \ref{fig:reject_example} shows examples of two sources that were rejected based on this visual inspection. These selection criteria are summarized in Table~\ref{tab:selection}. Given that only a small fraction of the targeted sources meet all requirements, it is clear that the criteria used for identifying KL sources from photometric catalogs was not ideal and needs further optimization. We discuss this issue along with potential improvements in Sec. \ref{sec:conclusions}.

We identify three source galaxies for which the rotation curve has flattened and passes visual examination, we refer to them as: b007, b008 and c007. Here, the leading letter denotes the slit mask and the following three digits indicate the slit number on that mask as assigned by the DEIMOS reduction pipeline. The attributes for these three source galaxies are listed in Table \ref{tab:KL_final_sample}.

\subsection{Results}

\begin{figure*}
    \centering
    \includegraphics[width=\linewidth,height=0.95\textheight,keepaspectratio]{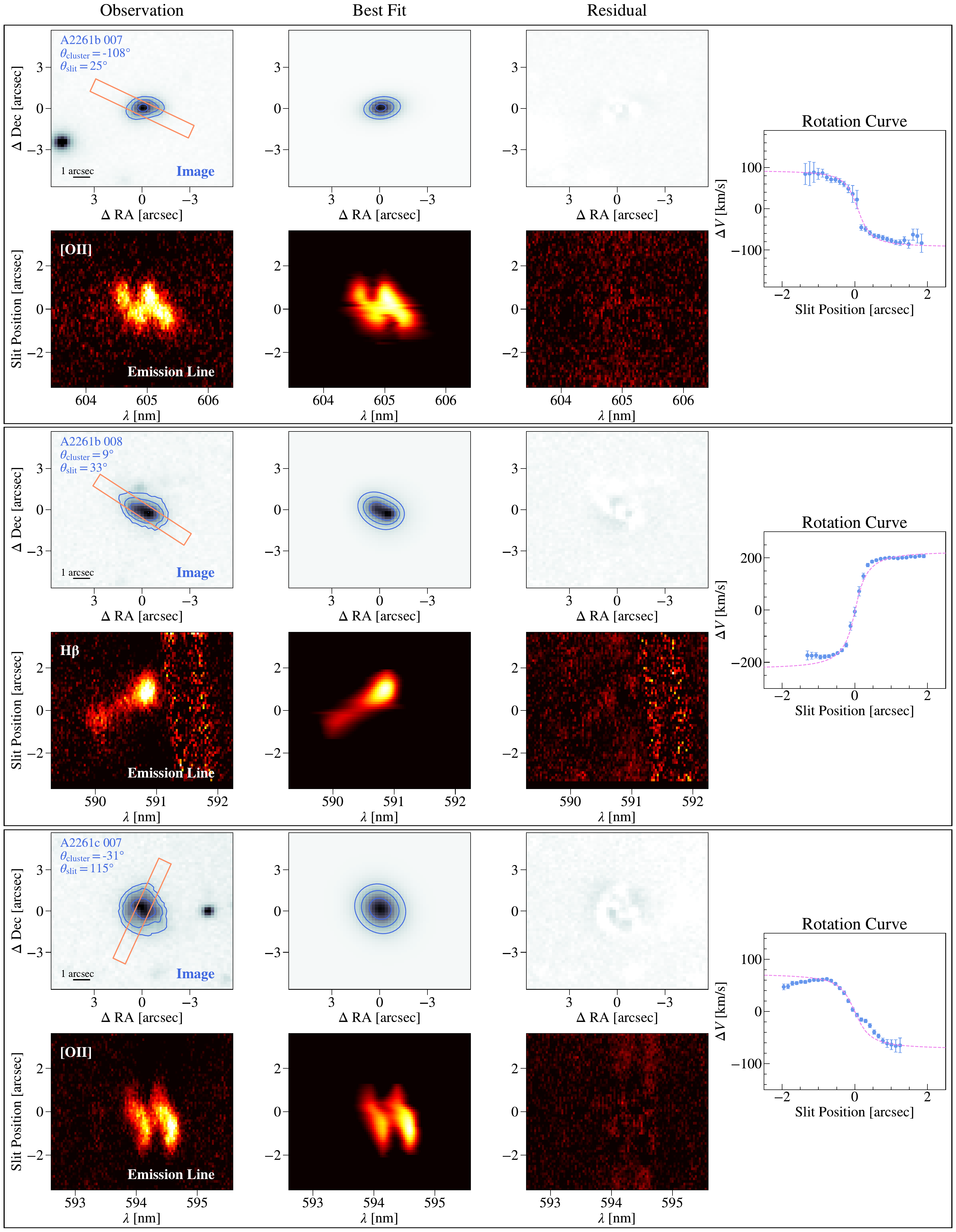}
    \caption{Best fit image and spectrum from the KL inference pipeline. The top, middle and bottom panels correspond to b007, b008 and c007, respectively. In each panel we show the observed Subaru $R_\mathrm{C}$-band image and DEIMOS spectrum along with best fits and residuals. We also show the rotation curve (w.r.t. the kinematic center) estimated from the observed emission line (blue circles) and the prediction from the best fit (pink dashed line).}
    \label{fig:bestfit_dv}
\end{figure*}

In Fig. \ref{fig:bestfit_dv}, the right-most panels show the rotation curves for our sources sample. For sources b007 and b008, we detect flattening of the rotation curve on both sides while c007 exhibits marginal flattening on one side only.

We apply the KL inference pipeline to the three sources and the best fit image and spectrum are shown in Fig. \ref{fig:bestfit_dv}. The tangential shear estimates around Abell 2261 are
\begin{itemize}
\item b007: $g_\mathrm{+,\, KL}=0.208\pm0.020$.
\item b008: $g_\mathrm{+,\, KL}=0.041\pm0.038$.
\item c007: $g_\mathrm{+,\, KL}=0.144\pm0.030$.
\end{itemize}

For each object, we measure a positive tangential shear, consistent with lensing by an overdense region. The measurement uncertainty averaged over our sample is $\sigma(g_+)\approx0.03$, which
is $\sim10\times$ better than the WL shape noise. These measurements demonstrate that KL can deliver high-precision shear measurements for individual sources even with spectra of moderate signal-to-noise. We simulate the three sources in their respective configurations, i.e. \{$\theta_\text{int}$, $\theta_\text{cluster}$, $\theta_\text{slit}$\}, and find that the variation in the shear measurement uncertainty is well predicted by the simulated analyses. We refer the reader to Appendix \ref{sec:mock_analyses} for more details.

\begin{figure*}
    \centering
    \includegraphics[width=\linewidth]{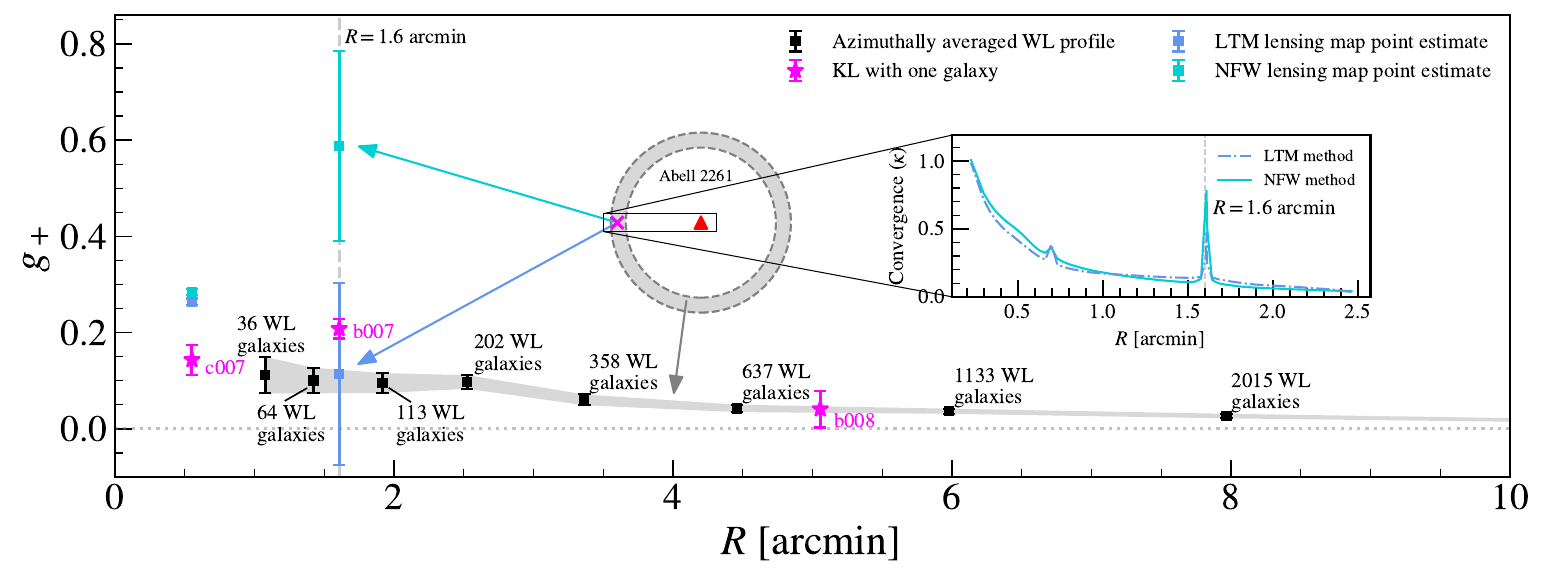}
    \caption{KL shear measurements from individual source galaxies (represented by magenta star symbols) compared to existing measurements in the literature. (i) The  shear profile of Abell 2261 is estimated using a total of 14510 WL galaxies distributed across 10 bins (8 shown here) \citep{Umetsu2014}. The galaxy counts in each bin are calculated assuming an average source density of 18.1 arcmin$^{-2}$. Note that the WL shear profile here is only suited for a qualitative comparison because of radial binning effects and rescaling to the average source redshift of the sample. For a more accurate comparison refer to Sec. \ref{sec:Clash_avg_profile}. (ii) The blue markers indicate the tangential shear measurement from the lensing map derived using a combined strong lensing and weak lensing analyses of HST observations \citep{Zitrin2015}. We find that b007 is located behind a substructure in the cluster. This is seen as a peak in the inset figure which shows a 1D slice of the convergence map in the direction of b007. As a result the shear prediction varies significantly over the region covered by the source when projected on the lensing map. The error bars on the blue markers represent  this systematic uncertainty which is computed from dispersion of shear prediction over the region covered by the source (as shown in Fig. \ref{fig:b007_cutout}).}
    \label{fig:kl_detect}
\end{figure*}

\subsection{Comparison with existing lensing measurements}
\label{sec:KL_comparison}
The KL shear measurements presented in the previous section align qualitatively with our expectations. We now shift our focus to a more quantitative comparison with shear measurements reported in the literature.

\subsubsection{Weak lensing measurements}
\label{sec:Clash_avg_profile}
\citet{Umetsu2014} performed a combined weak lensing shear and magnification analysis of the CLASH cluster sample using wide-field observations from Subaru Suprimecam and archival KPNO data. Figure \ref{fig:kl_detect} shows the azimuthally averaged tangential shear profile (black markers) measured from $0.9^\prime-16^\prime$ with an average source density of 18.1 arcmin$^{-2}$. The mean lensing depth for the CLASH WL source sample is $\langle\beta\rangle=0.7$, where the lensing depth is defined as
\begin{equation}
\label{eq:lensing_depth}
    \beta \equiv \frac{D_\mathrm{ls}}{D_\mathrm{s}}.
\end{equation}
Here, $D_\mathrm{s}$ is the angular diameter distance to the source, and $D_\mathrm{ls}$ is the angular diameter distance between the lens and source. In order to compare measurements, we multiply the average tangential shear estimate by $\langle g_+\rangle_\mathrm{WL} \times (\beta_\mathrm{KL}/\langle\beta\rangle$), where $\beta_\mathrm{KL}$ is the lensing depth for the KL sources (listed in Table \ref{tab:KL_final_sample}).

We predict the tangential shear at the radial separation of KL sources using the CLASH WL shape catalog
\begin{itemize}
    \item b007: $\langle g_+\rangle_\mathrm{WL}=0.13 \pm 0.025$ at $R=1.6^\prime$, using 113 sources b/w $R=[1.2^\prime, 2.0^\prime]$. 
    \item b008: $\langle g_+\rangle_\mathrm{WL}=0.059 \pm 0.012$ at $R=5.05^\prime$, using 532 sources b/w $R=[4.55^\prime, 5.55^\prime]$.
\end{itemize}
Note that as the WL shape catalog has no sources at $R<0.9^\prime$ we do not have a shear estimate for c007, which is at a separation of $R=0.55^\prime$.

For b007, the WL measurement using $\mathcal{O}(100)$ yields similar uncertainties as the KL measurement, demonstrating the precision of KL shear inference. We find that the KL measurement results in a shear approximately $1.6\times$ greater than the expectation from the WL shear profile. Since b007 is very close to the cluster center, this discrepancy is driven by the presence of cluster substructure as discussed in the next subsection.

The KL-inferred shear for b008 is in excellent agreement with the WL measurement. This is consistent with our expectation that KL and WL agree more as we go further away from the cluster center where there are fewer fluctuations in the underlying mass distribution.

\subsubsection{Strong lensing measurements}
The local tangential shear at a position around the cluster can deviate from the azimuthally averaged profile in the presence of lens substructure. We estimate the tangential shear at the source positions using the lensing maps from \citet{Zitrin2015}. The authors construct a mass model for Abell 2261 by combining \textit{Hubble Space Telescope} (HST) observations in the strong and weak lensing regimes. The mass models are constructed using two methods

\begin{enumerate}
    \item The Light Traces Mass (LTM) method \citep[see][for details]{Zitrin2009} assumes that the two main components of the mass model, galaxies and dark matter (DM), are traced by the cluster's light distribution. The galaxy component is modeled as a sum of contributions from cluster members, where each galaxy is assigned a luminosity weighted power law density profile. The DM contribution is the smoothed version of the galaxy component but added with a  relative weight. An external shear is applied to allow for additional flexibility and higher ellipticity of the critical curves.

    \item The PIEMDeNFW method (hereafter simply NFW) \citep[see][for details]{Zitrin2013} adopts the LTM assumption only for the galaxy component, however, in contrast to a power law each cluster member follows a luminosity scaled Pseudo-Isothermal Elliptical Mass Distribution (PIEMD). The DM component is modeled as an elliptical Navarro–Frenk–White profile \citep{Navarro1996} with mass, concentration, ellipticity, and position angle as free parameters. For complex and/or merging clusters an additional DM component is added.
\end{enumerate}

The mass models are constrained by simultaneously fitting 14 multiple-image systems and shapes of 718 WL sources, and then used to derive the weak lensing shears ($\gamma_1, \gamma_2$). As the regions covered by the lensing maps extend well into the non-linear regime, the reduced shear approximation might not be valid. We use the convergence map associated with the mass model to compute the reduced Cartesian shears as $g_{1,2}=\gamma_{1,2}/(1-\kappa)$. We note that the lensing and convergence maps are scaled to a source redshift such that the mean lensing depth $\beta=1$, therefore, we modify the shear and convergence estimated from the maps as  $\gamma_{1,2}\rightarrow\gamma_{1,2}\times\beta_\mathrm{KL}, \kappa\rightarrow\kappa\times\beta_\mathrm{KL}$. The tangential shear is calculated as per Eq. (\ref{eq:gplus}) at the location of the KL source \footnote{The lensing maps from \cite{Zitrin2015} are constructed assuming that the polar angle $\theta_\mathrm{cluster}$ is measured anti-clockwise from the $x-$axis, since we measure $\theta_\mathrm{cluster}$ clockwise from the RA axis we compute tangential shear from the lensing maps as $g_+ = -g_1\cos2\theta_\mathrm{cluster} + g_2 \sin2\theta_\mathrm{cluster}$.}.

While the mass models of \citet{Zitrin2015} incorporate WL data, their inference are predominantly driven by strong lensing constraints and modeling assumptions. As a result, their reported statistical errors significantly underestimate systematic uncertainties \citep[see][Sec. 5.1.4]{Zitrin2015}. We estimate the systematic uncertainty by computing the dispersion in the predicted tangential shear at the source position within a pixelized circular region, with the radius set to the best-fit disk half-light radius of the KL source. 

The shear predictions from LTM and NFW lensing maps along with the systematic uncertainties are
\begin{itemize}
    \item b007: 
    $g_\mathrm{+,LTM}=0.114\pm0.189$ and $g_\mathrm{+,NFW}=0.588\pm0.197$
    \item c007: $g_\mathrm{+,LTM}=0.264\pm0.007$ and  $g_\mathrm{+,NFW}=0.285\pm0.006$
\end{itemize}
These shear estimates are shown as blue markers in Fig. \ref{fig:kl_detect}. As the lensing maps only extend out to $R\approx3^\prime$ we do not have an estimate for b008 which is at $R=5.05^\prime$.

The KL shear for b007 is consistent with the prediction from the LTM method ($<$0.5$\sigma$) while discrepant with the NFW method  ($>$1.5$\sigma$) which predicts an unusually large shear. We find that this discrepancy is driven by the chance alignment of the source with substructure in the lens. We identify the substructure from a peak in a slice of the convergence map extending from the cluster center in the radial direction of b007 (inset plot in Fig. \ref{fig:kl_detect}). The two maps predict a convergence of $\kappa_\mathrm{LTM}\approx0.5,\kappa_\mathrm{NFW}\approx0.8$ at the source location. The impact of the substructure is also evident in the substantial difference between the lensing map predictions as shown in Fig. \ref{fig:b007_cutout}.

The KL shear measurement for c007 is smaller than lensing map predictions by more than $3\sigma$. Unlike b007, we do not find evidence for the presence of substructure near the source position. The two methods predict tangential shear consistent with one another, however, Fig. \ref{fig:shear_map_histogram} shows that this agreement is not representative of predictions at other locations, suggesting that the systematic uncertainty may still be underestimated. One other possible explanation for the observed discrepancy is irregularity in the source galaxy kinematics. The rotation curve for c007 (bottom panel of Fig. \ref{fig:bestfit_dv}), shows flattening only on one side and the comparison with the best-fit prediction suggests that the current model assumptions might not accurately capture the velocity field. Disk galaxies are known to exhibit structural features such as bars, warps, and spiral arms, which can locally distort the velocity field and lead to deviations from the assumption of symmetry \citep{Rix1995,Laine2014}. This intrinsic asymmetry induces a nonzero shear
$g_2^\mathrm{asymm}$, which results in a measured tangential shear of  $g_+= g_+^\mathrm{true} + g_2^\mathrm{asymm}\sin2\eta$, where $g_+^\mathrm{true}$ is the true tangential shear and $\eta$ is the angle between the tangential shear and source major axis. Depending on the source position relative to the lens, the induced shear can result in a positive or negative measurement bias\footnote{It is important to note that despite these potential astrophysical systematics, the average tangential shear estimated from randomly oriented sources in an annulus remains unbiased since $\langle\sin2\eta\rangle=0$.}. While making a definite statement is not possible without additional multi-slit/IFU observations, higher resolution HST images of the source reveal complex spiral arm structures which are not accounted for in the spectrum model.

It is important to exercise caution while interpreting the comparison with the shear derived from strong lensing mass models. First, strong lensing constraints extend out to approximately twice the Einstein radius \citep{Umetsu2016}. The effective Einstein radius for Abell 2261 is $R_\mathrm{ein}\approx23^{\prime\prime}$ (for source at $z=2$) \citep{Zitrin2015}, which suggests that the model predictions beyond $R\gtrapprox 40^{\prime\prime}$ are unreliable, as also evidenced by the huge difference in the mean shear predicted by the two methods for b007. Moreover, while the aperture mass contained within the critical curves can be determined with great accuracy and is nearly model-independent \citep{Oguri2012, Umetsu2016}, strong lensing does not offer a direct constraint on either shear or convergence. Although comparing different mass models can provide some insight into the impact of modeling choices, the differences between the two might underestimate the scatter since they share several assumptions. A more comprehensive understanding can be achieved by comparing a broader range of models such as those presented in the Frontier Fields lens models project \citep{Meneghetti2017}.

\section{Conclusions}
\label{sec:conclusions}
For the first time we use kinematic lensing to measure a cluster lensing signal. We select background disk galaxies from the CLASH catalog for spectroscopic follow-up with Keck DEIMOS. From the targets, we identify sources that can yield reliable shear measurements. The most important criteria for down-selection are the emission line signal-to-noise and unambiguous measurement of the flattening of the rotation curves. Out of the 141 targets, three sources meet the selection criteria.

We estimate the shear signal from the galaxy cluster Abell 2261 using three individual source galaxies with measurement uncertainty in the range $\sigma(g_+)=0.020-0.038$. The average shear uncertainty $\sigma(g_+)\approx0.03$ is equivalent to the precision achieved by stacking traditional WL measurements for $\sim100$ sources. These results are consistent with the forecasts presented in \citetalias{KLI} and  demonstrate that spectroscopic observations of moderate S/N are sufficient to reduce the shear measurement uncertainty by an order of magnitude.

We compare our shear estimates with measurements from the CLASH program. For the source b008, the KL shear is in good agreement with the average weak lensing shear profile while the other two sources (b007 and c007) also show broad consistency. We compare the KL shear for the latter two sources (located $<2^\prime$ from cluster center) with point estimates from strong lensing shear maps as the local shear can significantly deviate from the average shear close to the cluster center. We find that b007 is aligned with a substructure in the cluster which causes the LTM and NFW lensing maps to predict significantly different shears. Taking into account the uncertainty due to substructure in the shear maps, these predictions are consistent with our shear estimate. For c007, our shear estimate is in disagreement with predictions from the shear maps. Although we do not find nearby cluster substructure, there is evidence that the systematic uncertainty from assumptions in the strong lensing model is likely underestimated.

Part of the discrepancy for c007 can also be attributed to astrophysical systematics. Irregularities in galaxy kinematics are difficult to assess with single slit observations and can introduce biases in the shear inference. Ideally, the scatter in the Tully-Fisher relation should account for the range of kinematics expected from a galaxy population, and make KL robust against such systematics. However, in configurations where the KL shear inference primarily relies on kinematic misalignment (rather than the TF relation), an intrinsic asymmetry in the velocity field can be incorrectly interpreted as a shear signal. \citet{Gurri2020} identify similar trends, where this effect causes measurements that exceed theoretical predictions by a few sigma and even results in negative tangential shear in some instances.

We perform simulated analyses to explore the variation in the shear measurement uncertainty and find that it is strongly influenced by the lensing angle. Defined as the relative angle between the galaxy major axis and the tangential shear, the lensing angle can lead to a $2\times$ increase in uncertainty and in certain configurations the shear can be practically unconstrained. This variability arises because the lensing angle modulates the relative strength of $g_1^\mathrm{gal}$ and $g_2^\mathrm{gal}$ (defined in Eqs. \ref{eq:g1_gal} \& \ref{eq:g2_gal}),  hence in some configurations the tangential shear is constrained by the kinematic misalignment while in others the information primarily comes from the TF relation based prior. The latter provides a weaker constraint due to intrinsic scatter in the relation. Upon repeating the simulated analysis for configurations of the three KL sources, we find that the variation in observed shear uncertainties are within expectations.

The fact that we only find 3 usable galaxies out of the 141 targets implies that the observing strategy of the 2014 observing program was not optimal for KL inference. During down-selection, most source galaxies are excluded when a S/N threshold is applied; this requirement for reliable shear inference only became apparent in our recent forecasts in \citetalias{KLI} and was not clear when planning the observations in 2014. Previous studies have found that longer exposure times of 6-8 hours (as opposed to $\sim1$ hour for our observations) ensure high S/N at all radii and reliable measurement of the rotation curve \citep{Miller2011}. 

One approach for improving the success rate is to simply increase the exposure time of spectroscopic observations, which also ensures the detection of the (flattened) rotation curve from the faint outskirts of galaxies. Additionally, the fluxes of nebular emission lines can be predicted from observed photometry \citep{Valentino2017,Baugh2022} and can thus be used to identify more promising targets, improving the overall yield. Observing the targets with multiple slits can help in scenarios where the rotation curve is not properly sampled due to chance alignments of the slit mask with substructure in the source galaxy. In addition, multi-slit observations will enable us to measure $g_\times$, which can serve as a null test for residual systematics akin to B-mode measurements in WL.

The primary modeling challenge for KL lies in developing robust methods to capture complex morphology and kinematics. These issues are well-recognized in literature and numerous methods have been discussed e.g. analytical frameworks for modeling non-axisymmetric shapes \citep{Peng2010} and generalized approach using basis functions \citep{shapelets, sersiclets}. Additionally, more flexible kinematic models have been developed and successfully applied to data \citep{Jozsa2007a, Joszsa2007b, Labini2023}. Datasets such as from the MaNGA survey \citep{Bundy2015} offer a valuable resource for creating mock KL observations. These will enable robust testing of the inference pipeline  and also quantify the impact of increased model flexibility on constraining power. We defer such studies to future work.

This work highlights opportunities for the cosmological applications of KL. As demonstrated, cluster lensing is one such area where it can be utilized to calibrate scaling relations with significantly fewer sources. With a high signal-to-noise per source, KL can provide constraints on substructures within galaxy clusters --- a task not achievable by azimuthally averaged WL measurements --- thereby complementing current methods in distinguishing dark matter models using the subhalo mass function \citep{Gilman2020, Nadler2021}. By extension, it can significantly boost the number of ultra-diffuse galaxies and dwarf galaxies with halo mass measurements; these systems serve as unique testing grounds for theories of galaxy formation, modified gravity, and dark matter \citep{Sifon2018, Leauthaud2020, Thornton2023}.  Furthermore, the data required for these applications can already be obtained using existing multi-object spectrographs.

KL not only offers high statistical power but also ensures robustness against key WL systematics like photometric redshift uncertainties, contamination from IA, and shape measurement uncertainties. In the future, datasets from wide-field spectroscopic surveys like \textit{Euclid}, \textit{Roman}, and DESI II \citep{DESI-II} will broaden the scope for KL and enable its use for cosmic shear measurements. These advantages combined with the findings presented in this work suggest that KL holds significant potential in the era of precision cosmology.

\section*{Acknowledgements}
We thank Keiichi Umetsu for providing CLASH WL shear estimates and discussions relating to the CLASH analyses. We also thank Andrew Robertson, Anja von der Linden, and Kevin Bundy  for helpful discussions. This work was supported by NASA ROSES ADAP 20-ADAP20-0158. E.K. and P.R.S. were supported in part by the David and Lucile Packard Foundation and an Alfred P. Sloan Research Fellowship. The analyses in this work were carried out using the High Performance Computing (HPC) resources supported by the University of Arizona Technology and Research Initiative Fund (TRIF), University Information Technology Services (UITS), and the office for Research, Innovation, and Impact (RDI) and maintained by the UA Research Technologies Department.

The mass models were constructed by \citet{Zitrin2009, Zitrin2015}, and obtained through the Hubble Space Telescope Archive, as a high-end science product of the CLASH program \citep{Postman2012}. 

The Legacy Surveys consist of three individual and complementary projects: the Dark Energy Camera Legacy Survey (DECaLS; Proposal ID \#2014B-0404; PIs: David Schlegel and Arjun Dey), the Beijing-Arizona Sky Survey (BASS; NOAO Prop. ID \#2015A-0801; PIs: Zhou Xu and Xiaohui Fan), and the Mayall z-band Legacy Survey (MzLS; Prop. ID \#2016A-0453; PI: Arjun Dey). DECaLS, BASS and MzLS together include data obtained, respectively, at the Blanco telescope, Cerro Tololo Inter-American Observatory, NSF's NOIRLab; the Bok telescope, Steward Observatory, University of Arizona; and the Mayall telescope, Kitt Peak National Observatory, NOIRLab. Pipeline processing and analyses of the data were supported by NOIRLab and the Lawrence Berkeley National Laboratory (LBNL). The Legacy Surveys project is honored to be permitted to conduct astronomical research on Iolkam Du'ag (Kitt Peak), a mountain with particular significance to the Tohono O'odham Nation. NOIRLab is operated by the Association of Universities for Research in Astronomy (AURA) under a cooperative agreement with the National Science Foundation. LBNL is managed by the Regents of the University of California under contract to the U.S. Department of Energy. This project used data obtained with the Dark Energy Camera (DECam), which was constructed by the Dark Energy Survey (DES) collaboration. Funding for the DES Projects has been provided by the U.S. Department of Energy, the U.S. National Science Foundation, the Ministry of Science and Education of Spain, the Science and Technology Facilities Council of the United Kingdom, the Higher Education Funding Council for England, the National Center for Supercomputing Applications at the University of Illinois at Urbana-Champaign, the Kavli Institute of Cosmological Physics at the University of Chicago, Center for Cosmology and Astro-Particle Physics at the Ohio State University, the Mitchell Institute for Fundamental Physics and Astronomy at Texas A\&M University, Financiadora de Estudos e Projetos, Fundacao Carlos Chagas Filho de Amparo, Financiadora de Estudos e Projetos, Fundacao Carlos Chagas Filho de Amparo a Pesquisa do Estado do Rio de Janeiro, Conselho Nacional de Desenvolvimento Cientifico e Tecnologico and the Ministerio da Ciencia, Tecnologia e Inovacao, the Deutsche Forschungsgemeinschaft and the Collaborating Institutions in the Dark Energy Survey. The Collaborating Institutions are Argonne National Laboratory, the University of California at Santa Cruz, the University of Cambridge, Centro de Investigaciones Energeticas, Medioambientales y Tecnologicas-Madrid, the University of Chicago, University College London, the DES-Brazil Consortium, the University of Edinburgh, the Eidgenossische Technische Hochschule (ETH) Zurich, Fermi National Accelerator Laboratory, the University of Illinois at Urbana-Champaign, the Institut de Ciencies de l'Espai (IEEC/CSIC), the Institut de Fisica d'Altes Energies, Lawrence Berkeley National Laboratory, the Ludwig Maximilians Universitat Munchen and the associated Excellence Cluster Universe, the University of Michigan, NSF’s NOIRLab, the University of Nottingham, the Ohio State University, the University of Pennsylvania, the University of Portsmouth, SLAC National Accelerator Laboratory, Stanford University, the University of Sussex, and Texas A\&M University. BASS is a key project of the Telescope Access Program (TAP), which has been funded by the National Astronomical Observatories of China, the Chinese Academy of Sciences (the Strategic Priority Research Program “The Emergence of Cosmological Structures” Grant \# XDB09000000), and the Special Fund for Astronomy from the Ministry of Finance. The BASS is also supported by the External Cooperation Program of Chinese Academy of Sciences (Grant \# 114A11KYSB20160057), and Chinese National Natural Science Foundation (Grant \# 12120101003, \# 11433005). The Legacy Survey team makes use of data products from the Near-Earth Object Wide-field Infrared Survey Explorer (NEOWISE), which is a project of the Jet Propulsion Laboratory/California Institute of Technology. NEOWISE is funded by the National Aeronautics and Space Administration. The Legacy Surveys imaging of the DESI footprint is supported by the Director, Office of Science, Office of High Energy Physics of the U.S. Department of Energy under Contract No. DE-AC02-05CH1123, by the National Energy Research Scientific Computing Center, a DOE Office of Science User Facility under the same contract; and by the U.S. National Science Foundation, Division of Astronomical Sciences under Contract No. AST-0950945 to NOAO.
\section*{Data Availability}
The data underlying this article will be shared on reasonable request to the corresponding author.



\bibliographystyle{mnras}
\bibliography{bib} 

\begin{thebibliography}{}
\makeatletter
\relax
\def\mn@urlcharsother{\let\do\@makeother \do\$\do\&\do\#\do\^\do\_\do\%\do\~}
\def\mn@doi{\begingroup\mn@urlcharsother \@ifnextchar [ {\mn@doi@}
  {\mn@doi@[]}}
\def\mn@doi@[#1]#2{\def\@tempa{#1}\ifx\@tempa\@empty \href
  {http://dx.doi.org/#2} {doi:#2}\else \href {http://dx.doi.org/#2} {#1}\fi
  \endgroup}
\def\mn@eprint#1#2{\mn@eprint@#1:#2::\@nil}
\def\mn@eprint@arXiv#1{\href {http://arxiv.org/abs/#1} {{\tt arXiv:#1}}}
\def\mn@eprint@dblp#1{\href {http://dblp.uni-trier.de/rec/bibtex/#1.xml}
  {dblp:#1}}
\def\mn@eprint@#1:#2:#3:#4\@nil{\def\@tempa {#1}\def\@tempb {#2}\def\@tempc
  {#3}\ifx \@tempc \@empty \let \@tempc \@tempb \let \@tempb \@tempa \fi \ifx
  \@tempb \@empty \def\@tempb {arXiv}\fi \@ifundefined
  {mn@eprint@\@tempb}{\@tempb:\@tempc}{\expandafter \expandafter \csname
  mn@eprint@\@tempb\endcsname \expandafter{\@tempc}}}

\bibitem[\protect\citeauthoryear{{Amon} et~al.,}{{Amon}
  et~al.}{2022}]{Amon2022}
{Amon} A.,  et~al., 2022, \mn@doi [\prd] {10.1103/PhysRevD.105.023514}, \href
  {https://ui.adsabs.harvard.edu/abs/2022PhRvD.105b3514A} {105, 023514}

\bibitem[\protect\citeauthoryear{{Andrae}, {Melchior}  \& {Jahnke}}{{Andrae}
  et~al.}{2011}]{sersiclets}
{Andrae} R.,  {Melchior} P.,   {Jahnke} K.,  2011, \mn@doi [\mnras]
  {10.1111/j.1365-2966.2011.19348.x}, \href
  {https://ui.adsabs.harvard.edu/abs/2011MNRAS.417.2465A} {417, 2465}

\bibitem[\protect\citeauthoryear{{Asgari} et~al.,}{{Asgari}
  et~al.}{2021}]{Asgari2021}
{Asgari} M.,  et~al., 2021, \mn@doi [\aap] {10.1051/0004-6361/202039070}, \href
  {https://ui.adsabs.harvard.edu/abs/2021A&A...645A.104A} {645, A104}

\bibitem[\protect\citeauthoryear{{Baugh}, {Lacey}, {Gonzalez-Perez}  \&
  {Manzoni}}{{Baugh} et~al.}{2022}]{Baugh2022}
{Baugh} C.~M.,  {Lacey} C.~G.,  {Gonzalez-Perez} V.,   {Manzoni} G.,  2022,
  \mn@doi [\mnras] {10.1093/mnras/stab3506}, \href
  {https://ui.adsabs.harvard.edu/abs/2022MNRAS.510.1880B} {510, 1880}

\bibitem[\protect\citeauthoryear{{Ben{\'\i}tez}}{{Ben{\'\i}tez}}{2000}]{BPZ}
{Ben{\'\i}tez} N.,  2000, \mn@doi [\apj] {10.1086/308947}, \href
  {https://ui.adsabs.harvard.edu/abs/2000ApJ...536..571B} {536, 571}

\bibitem[\protect\citeauthoryear{{Bertin}}{{Bertin}}{2011}]{Bertin2011}
{Bertin} E.,  2011, in {Evans} I.~N.,  {Accomazzi} A.,  {Mink} D.~J.,   {Rots}
  A.~H.,  eds,  Astronomical Society of the Pacific Conference Series Vol. 442,
  Astronomical Data Analysis Software and Systems XX. p.~435

\bibitem[\protect\citeauthoryear{{Bertin} \& {Arnouts}}{{Bertin} \&
  {Arnouts}}{1996}]{Bertin1996}
{Bertin} E.,  {Arnouts} S.,  1996, \mn@doi [\aaps] {10.1051/aas:1996164}, \href
  {https://ui.adsabs.harvard.edu/abs/1996A&AS..117..393B} {117, 393}

\bibitem[\protect\citeauthoryear{{Blain}}{{Blain}}{2002}]{Blain2002}
{Blain} A.~W.,  2002, \mn@doi [\apjl] {10.1086/341103}, \href
  {https://ui.adsabs.harvard.edu/abs/2002ApJ...570L..51B} {570, L51}

\bibitem[\protect\citeauthoryear{{Bosch}}{{Bosch}}{2010}]{shapelets}
{Bosch} J.,  2010, \mn@doi [\aj] {10.1088/0004-6256/140/3/870}, \href
  {https://ui.adsabs.harvard.edu/abs/2010AJ....140..870B} {140, 870}

\bibitem[\protect\citeauthoryear{Buchner}{Buchner}{2014}]{Buchner2014}
Buchner J.,  2014, \mn@doi [Statistics and Computing]
  {10.1007/s11222-014-9512-y}, 26, 383–392

\bibitem[\protect\citeauthoryear{{Buchner}}{{Buchner}}{2019}]{Buchner2019}
{Buchner} J.,  2019, \mn@doi [\pasp] {10.1088/1538-3873/aae7fc}, \href
  {https://ui.adsabs.harvard.edu/abs/2019PASP..131j8005B} {131, 108005}

\bibitem[\protect\citeauthoryear{{Buchner}}{{Buchner}}{2021}]{buchner2021ultranest}
{Buchner} J.,  2021, \mn@doi [The Journal of Open Source Software]
  {10.21105/joss.03001}, \href
  {https://ui.adsabs.harvard.edu/abs/2021JOSS....6.3001B} {6, 3001}

\bibitem[\protect\citeauthoryear{{Bundy} et~al.,}{{Bundy}
  et~al.}{2015}]{Bundy2015}
{Bundy} K.,  et~al., 2015, \mn@doi [\apj] {10.1088/0004-637X/798/1/7}, \href
  {https://ui.adsabs.harvard.edu/abs/2015ApJ...798....7B} {798, 7}

\bibitem[\protect\citeauthoryear{{Chabrier}}{{Chabrier}}{2003}]{Chabrier2003}
{Chabrier} G.,  2003, \mn@doi [\pasp] {10.1086/376392}, \href
  {https://ui.adsabs.harvard.edu/abs/2003PASP..115..763C} {115, 763}

\bibitem[\protect\citeauthoryear{{Coil} et~al.,}{{Coil}
  et~al.}{2004}]{2004ApJ...609..525C}
{Coil} A.~L.,  et~al., 2004, \mn@doi [\apj] {10.1086/421337}, \href
  {https://ui.adsabs.harvard.edu/abs/2004ApJ...609..525C} {609, 525}

\bibitem[\protect\citeauthoryear{{Conroy} \& {Gunn}}{{Conroy} \&
  {Gunn}}{2010}]{Conroy2010}
{Conroy} C.,  {Gunn} J.~E.,  2010, \mn@doi [\apj]
  {10.1088/0004-637X/712/2/833}, \href
  {https://ui.adsabs.harvard.edu/abs/2010ApJ...712..833C} {712, 833}

\bibitem[\protect\citeauthoryear{{Conroy}, {Gunn}  \& {White}}{{Conroy}
  et~al.}{2009}]{Conroy2009}
{Conroy} C.,  {Gunn} J.~E.,   {White} M.,  2009, \mn@doi [\apj]
  {10.1088/0004-637X/699/1/486}, \href
  {https://ui.adsabs.harvard.edu/abs/2009ApJ...699..486C} {699, 486}

\bibitem[\protect\citeauthoryear{{Courteau}}{{Courteau}}{1997}]{Courteau1997}
{Courteau} S.,  1997, \mn@doi [\aj] {10.1086/118656}, \href
  {https://ui.adsabs.harvard.edu/abs/1997AJ....114.2402C} {114, 2402}

\bibitem[\protect\citeauthoryear{{Dark Energy Survey and Kilo-Degree Survey
  Collaboration}, {Abbott}  et~al.}{{Dark Energy Survey and Kilo-Degree Survey
  Collaboration} et~al.}{2023}]{Abott2023}
{Dark Energy Survey and Kilo-Degree Survey Collaboration} {Abbott} T.~M.~C.,
  et~al., 2023, \mn@doi [The Open Journal of Astrophysics]
  {10.21105/astro.2305.17173}, \href
  {https://ui.adsabs.harvard.edu/abs/2023OJAp....6E..36D} {6, 36}

\bibitem[\protect\citeauthoryear{{Davis} et~al.,}{{Davis}
  et~al.}{2003}]{2003SPIE.4834..161D}
{Davis} M.,  et~al., 2003, in {Guhathakurta} P.,  ed.,  Society of
  Photo-Optical Instrumentation Engineers (SPIE) Conference Series Vol. 4834,
  Discoveries and Research Prospects from 6- to 10-Meter-Class Telescopes II.
  pp 161--172 (\mn@eprint {arXiv} {astro-ph/0209419}),
  \mn@doi{10.1117/12.457897}

\bibitem[\protect\citeauthoryear{{Dey} et~al.,}{{Dey} et~al.}{2019}]{Dey2019}
{Dey} A.,  et~al., 2019, \mn@doi [\aj] {10.3847/1538-3881/ab089d}, \href
  {https://ui.adsabs.harvard.edu/abs/2019AJ....157..168D} {157, 168}

\bibitem[\protect\citeauthoryear{{DiGiorgio}, {Bundy}, {Westfall}, {Leauthaud}
  \& {Stark}}{{DiGiorgio} et~al.}{2021}]{DiGiorgio2021}
{DiGiorgio} B.,  {Bundy} K.,  {Westfall} K.~B.,  {Leauthaud} A.,   {Stark} D.,
  2021, \mn@doi [\apj] {10.3847/1538-4357/ac2572}, \href
  {https://ui.adsabs.harvard.edu/abs/2021ApJ...922..116D} {922, 116}

\bibitem[\protect\citeauthoryear{{Faber} et~al.,}{{Faber}
  et~al.}{2003}]{Faber2003}
{Faber} S.~M.,  et~al., 2003, in {Iye} M.,  {Moorwood} A. F.~M.,  eds,  Society
  of Photo-Optical Instrumentation Engineers (SPIE) Conference Series Vol.
  4841, Instrument Design and Performance for Optical/Infrared Ground-based
  Telescopes. pp 1657--1669, \mn@doi{10.1117/12.460346}

\bibitem[\protect\citeauthoryear{{Faber} et~al.,}{{Faber}
  et~al.}{2007}]{2007ApJ...665..265F}
{Faber} S.~M.,  et~al., 2007, \mn@doi [\apj] {10.1086/519294}, \href
  {https://ui.adsabs.harvard.edu/abs/2007ApJ...665..265F} {665, 265}

\bibitem[\protect\citeauthoryear{{Flaugher} et~al.,}{{Flaugher}
  et~al.}{2015}]{DECam}
{Flaugher} B.,  et~al., 2015, \mn@doi [\aj] {10.1088/0004-6256/150/5/150},
  \href {https://ui.adsabs.harvard.edu/abs/2015AJ....150..150F} {150, 150}

\bibitem[\protect\citeauthoryear{{Gilman}, {Birrer}, {Nierenberg}, {Treu}, {Du}
   \& {Benson}}{{Gilman} et~al.}{2020}]{Gilman2020}
{Gilman} D.,  {Birrer} S.,  {Nierenberg} A.,  {Treu} T.,  {Du} X.,   {Benson}
  A.,  2020, \mn@doi [\mnras] {10.1093/mnras/stz3480}, \href
  {https://ui.adsabs.harvard.edu/abs/2020MNRAS.491.6077G} {491, 6077}

\bibitem[\protect\citeauthoryear{{Grandis} et~al.,}{{Grandis}
  et~al.}{2024}]{Grandis2024}
{Grandis} S.,  et~al., 2024, \mn@doi [\aap] {10.1051/0004-6361/202348615},
  \href {https://ui.adsabs.harvard.edu/abs/2024A&A...687A.178G} {687, A178}

\bibitem[\protect\citeauthoryear{{Gurri}, {Taylor}  \& {Fluke}}{{Gurri}
  et~al.}{2020}]{Gurri2020}
{Gurri} P.,  {Taylor} E.~N.,   {Fluke} C.~J.,  2020, \mn@doi [\mnras]
  {10.1093/mnras/staa2893}, \href
  {https://ui.adsabs.harvard.edu/abs/2020MNRAS.499.4591G} {499, 4591}

\bibitem[\protect\citeauthoryear{{Huang}, {Krause}, {Xu}, {Eifler}, {R.~S.}  \&
  {Huff}}{{Huang} et~al.}{2024}]{Huang2024}
{Huang} Y.-H.,  {Krause} E.,  {Xu} J.,  {Eifler} T.,  {R.~S.} P.,   {Huff} E.,
  2024, \mn@doi [\prd] {10.1103/PhysRevD.110.043509}, \href
  {https://ui.adsabs.harvard.edu/abs/2024PhRvD.110d3509H} {110, 043509}

\bibitem[\protect\citeauthoryear{{Huff}, {Krause}, {Eifler}, {Fang}, {George}
  \& {Schlegel}}{{Huff} et~al.}{2013}]{Huff2013}
{Huff} E.~M.,  {Krause} E.,  {Eifler} T.,  {Fang} X.,  {George} M.~R.,
  {Schlegel} D.,  2013, \mn@doi [arXiv e-prints] {10.48550/arXiv.1311.1489},
  \href {https://ui.adsabs.harvard.edu/abs/2013arXiv1311.1489H} {p.
  arXiv:1311.1489}

\bibitem[\protect\citeauthoryear{{Ivezi{\'c}} et~al.,}{{Ivezi{\'c}}
  et~al.}{2019}]{LSST}
{Ivezi{\'c}} {\v Z}.,  et~al., 2019, \mn@doi [\apj] {10.3847/1538-4357/ab042c},
  \href {http://adsabs.harvard.edu/abs/2019ApJ...873..111I} {873, 111}

\bibitem[\protect\citeauthoryear{{Johnson}, {Leja}, {Conroy}  \&
  {Speagle}}{{Johnson} et~al.}{2021}]{Johnson2021}
{Johnson} B.~D.,  {Leja} J.,  {Conroy} C.,   {Speagle} J.~S.,  2021, \mn@doi
  [\apjs] {10.3847/1538-4365/abef67}, \href
  {https://ui.adsabs.harvard.edu/abs/2021ApJS..254...22J} {254, 22}

\bibitem[\protect\citeauthoryear{{J{\'o}zsa}}{{J{\'o}zsa}}{2007}]{Joszsa2007b}
{J{\'o}zsa} G.~I.~G.,  2007, \mn@doi [\aap] {10.1051/0004-6361:20066165}, \href
  {https://ui.adsabs.harvard.edu/abs/2007A&A...468..903J} {468, 903}

\bibitem[\protect\citeauthoryear{{J{\'o}zsa}, {Kenn}, {Klein}  \&
  {Oosterloo}}{{J{\'o}zsa} et~al.}{2007}]{Jozsa2007a}
{J{\'o}zsa} G.~I.~G.,  {Kenn} F.,  {Klein} U.,   {Oosterloo} T.~A.,  2007,
  \mn@doi [\aap] {10.1051/0004-6361:20066164}, \href
  {https://ui.adsabs.harvard.edu/abs/2007A&A...468..731J} {468, 731}

\bibitem[\protect\citeauthoryear{{Klein}, {Israel}, {Nagarajan}, {Bertoldi},
  {Pacaud}, {Lee}, {Sommer}  \& {Basu}}{{Klein} et~al.}{2019}]{Klein2019}
{Klein} M.,  {Israel} H.,  {Nagarajan} A.,  {Bertoldi} F.,  {Pacaud} F.,  {Lee}
  A.~T.,  {Sommer} M.,   {Basu} K.,  2019, \mn@doi [\mnras]
  {10.1093/mnras/stz1491}, \href
  {https://ui.adsabs.harvard.edu/abs/2019MNRAS.488.1704K} {488, 1704}

\bibitem[\protect\citeauthoryear{{Kriek} \& {Conroy}}{{Kriek} \&
  {Conroy}}{2013}]{Kriek2013}
{Kriek} M.,  {Conroy} C.,  2013, \mn@doi [\apjl] {10.1088/2041-8205/775/1/L16},
  \href {https://ui.adsabs.harvard.edu/abs/2013ApJ...775L..16K} {775, L16}

\bibitem[\protect\citeauthoryear{{Labini}, {Straccamore}, {De Marzo}  \&
  {Comer{\'o}n}}{{Labini} et~al.}{2023}]{Labini2023}
{Labini} F.~S.,  {Straccamore} M.,  {De Marzo} G.,   {Comer{\'o}n} S.,  2023,
  \mn@doi [\mnras] {10.1093/mnras/stad1916}, \href
  {https://ui.adsabs.harvard.edu/abs/2023MNRAS.524.1560S} {524, 1560}

\bibitem[\protect\citeauthoryear{{Laine} et~al.,}{{Laine}
  et~al.}{2014}]{Laine2014}
{Laine} S.,  et~al., 2014, \mn@doi [\mnras] {10.1093/mnras/stu1642}, \href
  {https://ui.adsabs.harvard.edu/abs/2014MNRAS.444.3015L} {444, 3015}

\bibitem[\protect\citeauthoryear{{Lang}, {Hogg}  \& {Mykytyn}}{{Lang}
  et~al.}{2016}]{Lang2016}
{Lang} D.,  {Hogg} D.~W.,   {Mykytyn} D.,  2016, {The Tractor: Probabilistic
  astronomical source detection and measurement}, Astrophysics Source Code
  Library, record ascl:1604.008

\bibitem[\protect\citeauthoryear{{Laureijs} et~al.,}{{Laureijs}
  et~al.}{2011}]{Euclid}
{Laureijs} R.,  et~al., 2011, arXiv e-prints, \href
  {https://ui.adsabs.harvard.edu/abs/2011arXiv1110.3193L} {p. arXiv:1110.3193}

\bibitem[\protect\citeauthoryear{{Leauthaud}, {Singh}, {Luo}, {Ardila},
  {Greco}, {Capak}, {Greene}  \& {Mayer}}{{Leauthaud}
  et~al.}{2020}]{Leauthaud2020}
{Leauthaud} A.,  {Singh} S.,  {Luo} Y.,  {Ardila} F.,  {Greco} J.~P.,  {Capak}
  P.,  {Greene} J.~E.,   {Mayer} L.,  2020, \mn@doi [Physics of the Dark
  Universe] {10.1016/j.dark.2020.100719}, \href
  {https://ui.adsabs.harvard.edu/abs/2020PDU....3000719L} {30, 100719}

\bibitem[\protect\citeauthoryear{{Li} et~al.,}{{Li} et~al.}{2023}]{Li2023}
{Li} X.,  et~al., 2023, \mn@doi [\prd] {10.1103/PhysRevD.108.123518}, \href
  {https://ui.adsabs.harvard.edu/abs/2023PhRvD.108l3518L} {108, 123518}

\bibitem[\protect\citeauthoryear{{Meneghetti} et~al.,}{{Meneghetti}
  et~al.}{2017}]{Meneghetti2017}
{Meneghetti} M.,  et~al., 2017, \mn@doi [\mnras] {10.1093/mnras/stx2064}, \href
  {https://ui.adsabs.harvard.edu/abs/2017MNRAS.472.3177M} {472, 3177}

\bibitem[\protect\citeauthoryear{{Miller}, {Bundy}, {Sullivan}, {Ellis}  \&
  {Treu}}{{Miller} et~al.}{2011}]{Miller2011}
{Miller} S.~H.,  {Bundy} K.,  {Sullivan} M.,  {Ellis} R.~S.,   {Treu} T.,
  2011, \mn@doi [\apj] {10.1088/0004-637X/741/2/115}, \href
  {https://ui.adsabs.harvard.edu/abs/2011ApJ...741..115M} {741, 115}

\bibitem[\protect\citeauthoryear{{Molino} et~al.,}{{Molino}
  et~al.}{2017}]{2017MNRAS.470...95M}
{Molino} A.,  et~al., 2017, \mn@doi [\mnras] {10.1093/mnras/stx1243}, \href
  {https://ui.adsabs.harvard.edu/abs/2017MNRAS.470...95M} {470, 95}

\bibitem[\protect\citeauthoryear{{Morales}}{{Morales}}{2006}]{Morales2006}
{Morales} M.~F.,  2006, \mn@doi [\apjl] {10.1086/508614}, \href
  {https://ui.adsabs.harvard.edu/abs/2006ApJ...650L..21M} {650, L21}

\bibitem[\protect\citeauthoryear{{Murray}, {Bartlett}, {Artis}  \&
  {Melin}}{{Murray} et~al.}{2022}]{Murray2022}
{Murray} C.,  {Bartlett} J.~G.,  {Artis} E.,   {Melin} J.-B.,  2022, \mn@doi
  [\mnras] {10.1093/mnras/stac689}, \href
  {https://ui.adsabs.harvard.edu/abs/2022MNRAS.512.4785M} {512, 4785}

\bibitem[\protect\citeauthoryear{{Nadler}, {Birrer}, {Gilman}, {Wechsler},
  {Du}, {Benson}, {Nierenberg}  \& {Treu}}{{Nadler} et~al.}{2021}]{Nadler2021}
{Nadler} E.~O.,  {Birrer} S.,  {Gilman} D.,  {Wechsler} R.~H.,  {Du} X.,
  {Benson} A.,  {Nierenberg} A.~M.,   {Treu} T.,  2021, \mn@doi [\apj]
  {10.3847/1538-4357/abf9a3}, \href
  {https://ui.adsabs.harvard.edu/abs/2021ApJ...917....7N} {917, 7}

\bibitem[\protect\citeauthoryear{{Navarro}, {Frenk}  \& {White}}{{Navarro}
  et~al.}{1996}]{Navarro1996}
{Navarro} J.~F.,  {Frenk} C.~S.,   {White} S. D.~M.,  1996, \mn@doi [\apj]
  {10.1086/177173}, \href
  {https://ui.adsabs.harvard.edu/abs/1996ApJ...462..563N} {462, 563}

\bibitem[\protect\citeauthoryear{{Newman} et~al.,}{{Newman}
  et~al.}{2013}]{deep2}
{Newman} J.~A.,  et~al., 2013, \mn@doi [\apjs] {10.1088/0067-0049/208/1/5},
  \href {https://ui.adsabs.harvard.edu/abs/2013ApJS..208....5N} {208, 5}

\bibitem[\protect\citeauthoryear{{Oguri}, {Bayliss}, {Dahle}, {Sharon},
  {Gladders}, {Natarajan}, {Hennawi}  \& {Koester}}{{Oguri}
  et~al.}{2012}]{Oguri2012}
{Oguri} M.,  {Bayliss} M.~B.,  {Dahle} H.,  {Sharon} K.,  {Gladders} M.~D.,
  {Natarajan} P.,  {Hennawi} J.~F.,   {Koester} B.~P.,  2012, \mn@doi [\mnras]
  {10.1111/j.1365-2966.2011.20248.x}, \href
  {https://ui.adsabs.harvard.edu/abs/2012MNRAS.420.3213O} {420, 3213}

\bibitem[\protect\citeauthoryear{{Okabe} \& {Smith}}{{Okabe} \&
  {Smith}}{2016}]{Okabe2016}
{Okabe} N.,  {Smith} G.~P.,  2016, \mn@doi [\mnras] {10.1093/mnras/stw1539},
  \href {https://ui.adsabs.harvard.edu/abs/2016MNRAS.461.3794O} {461, 3794}

\bibitem[\protect\citeauthoryear{{Peng}, {Ho}, {Impey}  \& {Rix}}{{Peng}
  et~al.}{2010}]{Peng2010}
{Peng} C.~Y.,  {Ho} L.~C.,  {Impey} C.~D.,   {Rix} H.-W.,  2010, \mn@doi [\aj]
  {10.1088/0004-6256/139/6/2097}, \href
  {https://ui.adsabs.harvard.edu/abs/2010AJ....139.2097P} {139, 2097}

\bibitem[\protect\citeauthoryear{{Persic}, {Salucci}  \& {Stel}}{{Persic}
  et~al.}{1996}]{Persic1996}
{Persic} M.,  {Salucci} P.,   {Stel} F.,  1996, \mn@doi [\mnras]
  {10.1093/mnras/278.1.27}, \href
  {https://ui.adsabs.harvard.edu/abs/1996MNRAS.281...27P} {281, 27}

\bibitem[\protect\citeauthoryear{{Postman} et~al.,}{{Postman}
  et~al.}{2012}]{Postman2012}
{Postman} M.,  et~al., 2012, \mn@doi [\apjs] {10.1088/0067-0049/199/2/25},
  \href {https://ui.adsabs.harvard.edu/abs/2012ApJS..199...25P} {199, 25}

\bibitem[\protect\citeauthoryear{{R.~S.}, Krause, Huang, Huff, Xu, Eifler  \&
  Everett}{{R.~S.} et~al.}{2023}]{KLI}
{R.~S.} P.,  Krause E.,  Huang H.-J.,  Huff E.,  Xu J.,  Eifler T.,   Everett
  S.,  2023, \mn@doi [Monthly Notices of the Royal Astronomical Society]
  {10.1093/mnras/stad2014}, 524, 3324

\bibitem[\protect\citeauthoryear{{Rix} \& {Zaritsky}}{{Rix} \&
  {Zaritsky}}{1995}]{Rix1995}
{Rix} H.-W.,  {Zaritsky} D.,  1995, \mn@doi [\apj] {10.1086/175858}, \href
  {https://ui.adsabs.harvard.edu/abs/1995ApJ...447...82R} {447, 82}

\bibitem[\protect\citeauthoryear{{Rowe} et~al.,}{{Rowe} et~al.}{2015}]{RJM+15}
{Rowe} B.~T.~P.,  et~al., 2015, \mn@doi [Astronomy and Computing]
  {10.1016/j.ascom.2015.02.002}, \href
  {https://ui.adsabs.harvard.edu/abs/2015A&C....10..121R} {10, 121}

\bibitem[\protect\citeauthoryear{{Schlegel} et~al.,}{{Schlegel}
  et~al.}{2022}]{DESI-II}
{Schlegel} D.~J.,  et~al., 2022, arXiv e-prints, \href
  {https://ui.adsabs.harvard.edu/abs/2022arXiv220903585S} {p. arXiv:2209.03585}

\bibitem[\protect\citeauthoryear{{Schmidt} et~al.,}{{Schmidt}
  et~al.}{2020}]{Schmidt2020}
{Schmidt} S.~J.,  et~al., 2020, \mn@doi [\mnras] {10.1093/mnras/staa2799},
  \href {https://ui.adsabs.harvard.edu/abs/2020MNRAS.499.1587S} {499, 1587}

\bibitem[\protect\citeauthoryear{{Sif{\'o}n}, {van der Burg}, {Hoekstra},
  {Muzzin}  \& {Herbonnet}}{{Sif{\'o}n} et~al.}{2018}]{Sifon2018}
{Sif{\'o}n} C.,  {van der Burg} R. F.~J.,  {Hoekstra} H.,  {Muzzin} A.,
  {Herbonnet} R.,  2018, \mn@doi [\mnras] {10.1093/mnras/stx2648}, \href
  {https://ui.adsabs.harvard.edu/abs/2018MNRAS.473.3747S} {473, 3747}

\bibitem[\protect\citeauthoryear{{Spergel} et~al.,}{{Spergel}
  et~al.}{2015}]{Roman}
{Spergel} D.,  et~al., 2015, arXiv e-prints, \href
  {https://ui.adsabs.harvard.edu/abs/2015arXiv150303757S} {p. arXiv:1503.03757}

\bibitem[\protect\citeauthoryear{{Thornton} et~al.,}{{Thornton}
  et~al.}{2023}]{Thornton2023}
{Thornton} J.,  et~al., 2023, \mn@doi [arXiv e-prints]
  {10.48550/arXiv.2311.14659}, \href
  {https://ui.adsabs.harvard.edu/abs/2023arXiv231114659T} {p. arXiv:2311.14659}

\bibitem[\protect\citeauthoryear{{Tully} \& {Fisher}}{{Tully} \&
  {Fisher}}{1977}]{Tully1977}
{Tully} R.~B.,  {Fisher} J.~R.,  1977, \aap, \href
  {https://ui.adsabs.harvard.edu/abs/1977A&A....54..661T} {54, 661}

\bibitem[\protect\citeauthoryear{{Umetsu}}{{Umetsu}}{2020}]{Umetsu2020}
{Umetsu} K.,  2020, \mn@doi [\aapr] {10.1007/s00159-020-00129-w}, \href
  {https://ui.adsabs.harvard.edu/abs/2020A&ARv..28....7U} {28, 7}

\bibitem[\protect\citeauthoryear{{Umetsu} et~al.,}{{Umetsu}
  et~al.}{2014}]{Umetsu2014}
{Umetsu} K.,  et~al., 2014, \mn@doi [\apj] {10.1088/0004-637X/795/2/163}, \href
  {https://ui.adsabs.harvard.edu/abs/2014ApJ...795..163U} {795, 163}

\bibitem[\protect\citeauthoryear{{Umetsu}, {Zitrin}, {Gruen}, {Merten},
  {Donahue}  \& {Postman}}{{Umetsu} et~al.}{2016}]{Umetsu2016}
{Umetsu} K.,  {Zitrin} A.,  {Gruen} D.,  {Merten} J.,  {Donahue} M.,
  {Postman} M.,  2016, \mn@doi [\apj] {10.3847/0004-637X/821/2/116}, \href
  {https://ui.adsabs.harvard.edu/abs/2016ApJ...821..116U} {821, 116}

\bibitem[\protect\citeauthoryear{{Valentino} et~al.,}{{Valentino}
  et~al.}{2017}]{Valentino2017}
{Valentino} F.,  et~al., 2017, \mn@doi [\mnras] {10.1093/mnras/stx2305}, \href
  {https://ui.adsabs.harvard.edu/abs/2017MNRAS.472.4878V} {472, 4878}

\bibitem[\protect\citeauthoryear{{Wright} et~al.,}{{Wright}
  et~al.}{2010}]{WISE}
{Wright} E.~L.,  et~al., 2010, \mn@doi [\aj] {10.1088/0004-6256/140/6/1868},
  \href {https://ui.adsabs.harvard.edu/abs/2010AJ....140.1868W} {140, 1868}

\bibitem[\protect\citeauthoryear{{Xu}, {Eifler}, {Huff}, {Pranjal}, {Huang},
  {Everett}  \& {Krause}}{{Xu} et~al.}{2023}]{Xu2023}
{Xu} J.,  {Eifler} T.,  {Huff} E.,  {Pranjal} R.~S.,  {Huang} H.-J.,  {Everett}
  S.,   {Krause} E.,  2023, \mn@doi [\mnras] {10.1093/mnras/stac3685}, \href
  {https://ui.adsabs.harvard.edu/abs/2023MNRAS.519.2535X} {519, 2535}

\bibitem[\protect\citeauthoryear{{Xu}, {Eifler}, {Wang}, {Krause}, {Everett},
  {Huff}, {Pranjal R.}  \& {Huang}}{{Xu} et~al.}{2024}]{Xu2024}
{Xu} J.,  {Eifler} T.,  {Wang} E.,  {Krause} E.,  {Everett} S.,  {Huff} E.,
  {Pranjal R.} S.,   {Huang} Y.-H.,  2024, \mn@doi [arXiv e-prints]
  {10.48550/arXiv.2407.20867}, \href
  {https://ui.adsabs.harvard.edu/abs/2024arXiv240720867X} {p. arXiv:2407.20867}

\bibitem[\protect\citeauthoryear{{Zhang}, {Rau}, {Mandelbaum}, {Li}  \&
  {Moews}}{{Zhang} et~al.}{2023}]{Zhang2023}
{Zhang} T.,  {Rau} M.~M.,  {Mandelbaum} R.,  {Li} X.,   {Moews} B.,  2023,
  \mn@doi [\mnras] {10.1093/mnras/stac3090}, \href
  {https://ui.adsabs.harvard.edu/abs/2023MNRAS.518..709Z} {518, 709}

\bibitem[\protect\citeauthoryear{{Zitrin} et~al.,}{{Zitrin}
  et~al.}{2009}]{Zitrin2009}
{Zitrin} A.,  et~al., 2009, \mn@doi [\mnras]
  {10.1111/j.1365-2966.2009.14899.x}, \href
  {https://ui.adsabs.harvard.edu/abs/2009MNRAS.396.1985Z} {396, 1985}

\bibitem[\protect\citeauthoryear{{Zitrin} et~al.,}{{Zitrin}
  et~al.}{2013}]{Zitrin2013}
{Zitrin} A.,  et~al., 2013, \mn@doi [\apjl] {10.1088/2041-8205/762/2/L30},
  \href {https://ui.adsabs.harvard.edu/abs/2013ApJ...762L..30Z} {762, L30}

\bibitem[\protect\citeauthoryear{{Zitrin} et~al.,}{{Zitrin}
  et~al.}{2015}]{Zitrin2015}
{Zitrin} A.,  et~al., 2015, \mn@doi [\apj] {10.1088/0004-637X/801/1/44}, \href
  {https://ui.adsabs.harvard.edu/abs/2015ApJ...801...44Z} {801, 44}

\bibitem[\protect\citeauthoryear{{de Burgh-Day}, {Taylor}, {Webster}  \&
  {Hopkins}}{{de Burgh-Day} et~al.}{2015}]{2015MNRAS.451.2161D}
{de Burgh-Day} C.~O.,  {Taylor} E.~N.,  {Webster} R.~L.,   {Hopkins} A.~M.,
  2015, \mn@doi [\mnras] {10.1093/mnras/stv1083}, \href
  {https://ui.adsabs.harvard.edu/abs/2015MNRAS.451.2161D} {451, 2161}

\bibitem[\protect\citeauthoryear{{von der Linden} et~al.,}{{von der Linden}
  et~al.}{2014}]{WtG2014a}
{von der Linden} A.,  et~al., 2014, \mn@doi [\mnras] {10.1093/mnras/stt1945},
  \href {https://ui.adsabs.harvard.edu/abs/2014MNRAS.439....2V} {439, 2}

\makeatother
\end{thebibliography}

\appendix
\section{Impact of source galaxy orientation on shear uncertainty}
\label{sec:mock_analyses}
\begin{figure*}[H]
    \centering
    \includegraphics[width=\linewidth]{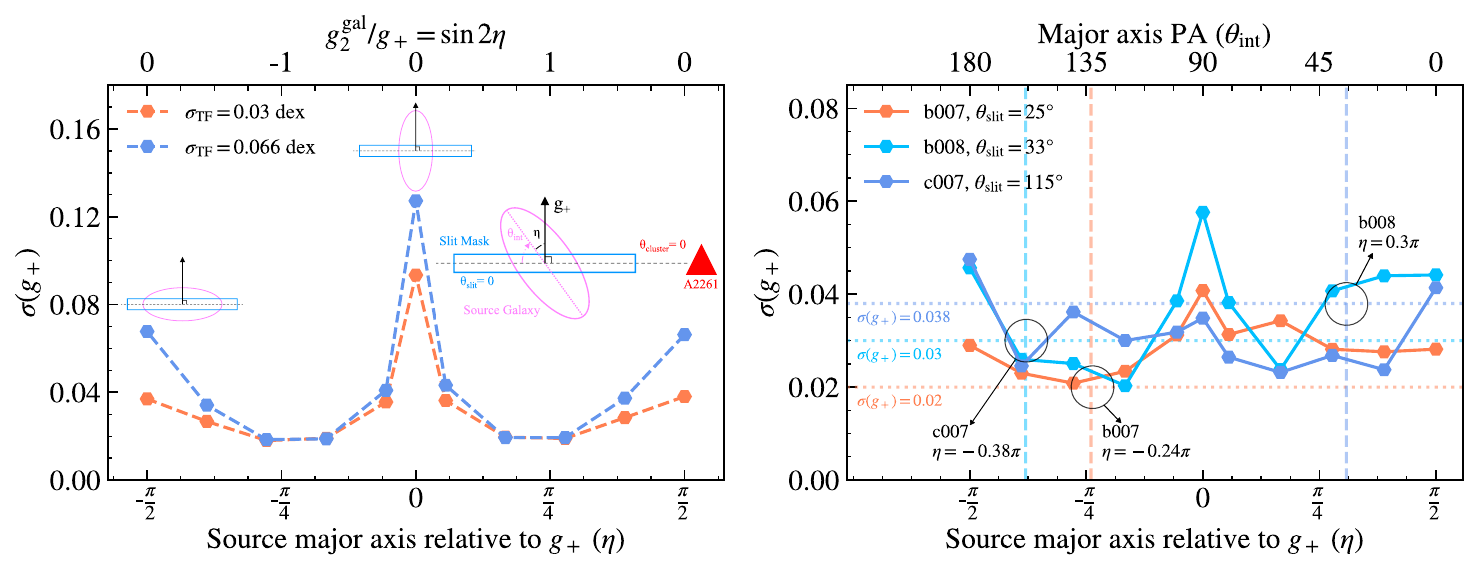}
    \caption{Simulating the impact of galaxy orientation on shear measurement uncertainty. Left: The galaxy position is fixed at $\theta_\mathrm{cluster}=0$, meaning that the tangential shear always points in the $y$-direction while we vary $\theta_\mathrm{int}$ as illustrated in the schematic. Kinematic misalignment provides a better constraint on the tangential shear than the TF relation, hence $\sigma(g_+)$ is smallest when $g_2^\mathrm{gal}\gtrapprox g_1^\mathrm{gal}$. This is also corroborated by the fact the tightening the TF prior improves $\sigma(g_+)$ the most when $|\eta|=0,\pi/2$. Right: Same as left panel but using slit PA and inclination specific to the three KL sources. The vertical and horizontal lines indicate the lensing angle and shear uncertainty from data, respectively. The simulated analysis approximately reproduces the shear uncertainty from the data (indicated by black circles) and demonstrates that the variation in $\sigma(g_+)$ among the sources can be attributed to differences in slit orientation and lensing angle.} 
    \label{fig:shear_err_vs_eta}
\end{figure*}

The shear measurement uncertainty for the KL sample ranges from 0.02 to 0.038, which is approximately a twofold variation. Although the emission line S/N for b007 is slightly higher than b008, the analysis in \citetalias{KLI} suggests that the S/N would need to be over three times greater to account for the observed differences. Additionally, even though galaxies with lower inclination yield tighter shear constraints (see \citetalias{KLI}, Sec. 5.2), this factor is unlikely to explain the 50\% difference in measurement uncertainty between b007 and c007 as they have similar best-fit inclinations (approximately 25° and 30°, respectively).

One possible explanation is the relative orientation of the galaxy major axis and the tangential shear, which we refer to as the lensing angle $\eta$. The tangential shear $g_+$ can be transformed into the galaxy reference frame, where the major and minor axes align with the coordinate axes, as
\begin{align}
    g_1^\mathrm{gal}&=g_+\cos 2\eta \label{eq:g1_gal},\\
    g_2^\mathrm{gal}&=g_+\sin 2\eta \label{eq:g2_gal}.
\end{align}
As described in Sect. \ref{sec:kl_theory}, the KL technique estimates the $g_1$ component using a TF relation-based prior while the $g_2$ component is inferred from kinematic misalignment; the former is less constraining due to the astrophysical scatter in the TF relation. This implies that the lensing angle influences the measurement uncertainty.

We conduct simulated analyses to examine how shear uncertainty $\sigma(g_+)$ varies with the lensing angle. We generate noiseless mock data for a source galaxy at $\theta_\mathrm{cluster}=0$, setting the mock image and spectrum to $\mathrm{S/N}=72$ and 34, respectively (as in the case of b008). For simplicity, we fix the fiducial slit position angle ($\theta_\mathrm{slit}$) at zero and vary the intrinsic position angle ($\theta_\mathrm{int}$) in the range $(0,\pi)$, refer to Fig. \ref{fig:lensing_geometry} for a schematic of the lensing geometry. Note that the lensing angle is given by $\eta=\pi/2-\theta_\mathrm{int}$ when $\theta_\mathrm{cluster}=0$. 

As shown in the left panel of Fig. \ref{fig:shear_err_vs_eta}, the shear measurement uncertainty strongly depends on $\eta$. $\sigma(g_+)$ is largest when the major axis aligns with (or is perpendicular to) the tangential shear i.e. $|\eta|=0, \pi/2$ and smallest when $\eta\approx\pi/4$. This behavior is explained by the fact that when $g_2^\mathrm{gal}$ is zero (Eq. \ref{eq:g2_gal}), the velocity field remains symmetric, and the inference relies on the TF relation, resulting in a higher measurement uncertainty. This is also evidenced by the fact that the orientations with $|\eta|=0, \pi/2$ show the most improvement in $\sigma(g_+)$ when narrowing the TF relation based prior (going from blue to orange curve), while other configurations show marginal gains.

The relative orientation of the slit and the galaxy major axis also impacts the measurement, as seen from the difference in $\sigma(g_+)$ for $\eta=0$ and $|\eta|=\pi/2$. In both cases we rely entirely on the TF relation (since $g_2^\mathrm{gal}=0$), the only difference is that at $\eta=0$ we measure the kinematics along the minor axis while the slit is aligned along the major axis when $|\eta|=\pi/2$. The former configuration does not provide information about the maximum line-of-sight velocity, which is required to estimate the intrinsic shape in conjunction with the TF relation. This makes the shear practically unconstrained in the $\eta=0$ configuration.

To summarize, the kinematic misalignment induced by the $g_2$ component provides more information than the $g_1$ component, which relies on the TF relation. Therefore, the ability to constrain tangential shear depends on the lensing angle as it dictates the relative strength of the two shear components. The lensing angle also influences the slit angle that minimizes measurement uncertainty; optimal constraints for $g_1$ and $g_2$ are achieved when the slit is aligned with the major and minor axis, respectively. Deviations from this alignment generally reduce the constraining power.

With these considerations in mind, we can better understand the measurement uncertainty for the KL sources. In the right panel of Fig. \ref{fig:shear_err_vs_eta} we vary the lensing angle (by varying $\theta_\mathrm{int}$ and fixing $\theta_\mathrm{cluster}=0$) while adopting the slit configuration and best-fit inclinations of the three KL sources. In all cases, the simulated analysis  approximately matches the shear uncertainties from data (marked by the circles), we discuss these below. 
\begin{itemize}
    \item b007: The measurement uncertainty is the lowest among the three sources since (a) $\eta=-0.24\pi$ implying $|g_2^\mathrm{gal}/g_+|\approx1$ (b) the slit is closely aligned with the minor axis with an offset $\Delta \theta_\mathrm{minor}=\theta_\mathrm{{int}}-\theta_\mathrm{slit}-90\degree=133\degree-25\degree-90\degree=18\degree$.
    \item b008: Even though the lensing angle induces significant kinematic misalignment ($|g_2^\mathrm{gal}/g_+|\approx0.95$), the measurement uncertainty is larger than b007 due to the slit being aligned with the major axis $\Delta \theta_\mathrm{major}=\theta_\mathrm{{int}}-\theta_\mathrm{slit}=36\degree-33\degree=3\degree$.
    \item c007: In this case the induced kinematic misalignment ($|g_2^\mathrm{gal}/g_+|\approx0.68$) is less than b008,  but the measurement uncertainty is smaller owing to the slit orientation $\Delta \theta_\mathrm{major}=\theta_\mathrm{{int}}-\theta_\mathrm{slit}=158\degree-115\degree=43\degree$.
\end{itemize}
The remaining features in the figure can be similarly understood. These results suggest that the variation in measured shear uncertainties is a consequence of the interplay between galaxy orientation, lensing geometry and observing strategy.

\section{Comparison of strong lensing shear maps}

\begin{figure}
    \centering
    \includegraphics[width=\linewidth]{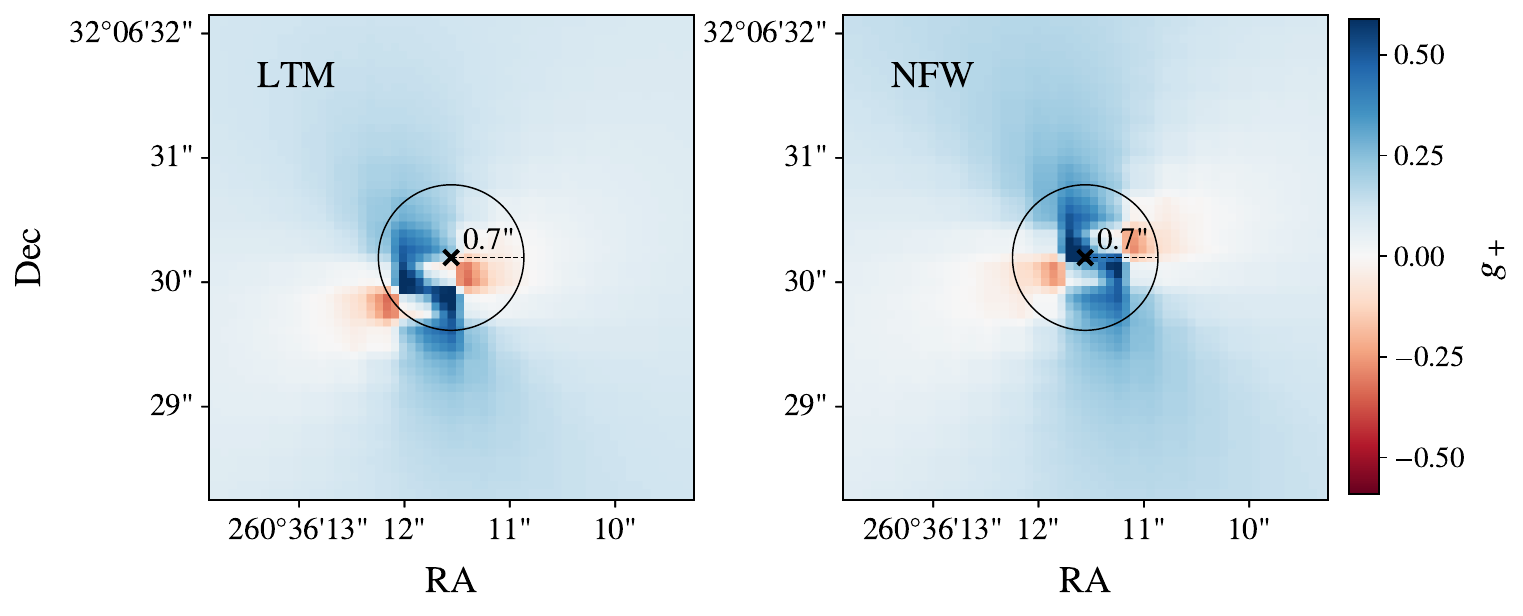}
    \caption{Cutouts from the tangential shear maps from the LTM and NFW methods. The cross symbol marks the location of b007. The shear estimate varies significantly across the region covered by the source galaxy due to a chance alignment with sub-structure in the foreground lens. We compute the dispersion in the lensing map shear over a circular region with radius set to the disk half-light radius, and find that the sub-structure results in an uncertainty of $\approx0.2$.}
    \label{fig:b007_cutout}
\end{figure}

\begin{figure}
    \centering
    \includegraphics[width=\linewidth]{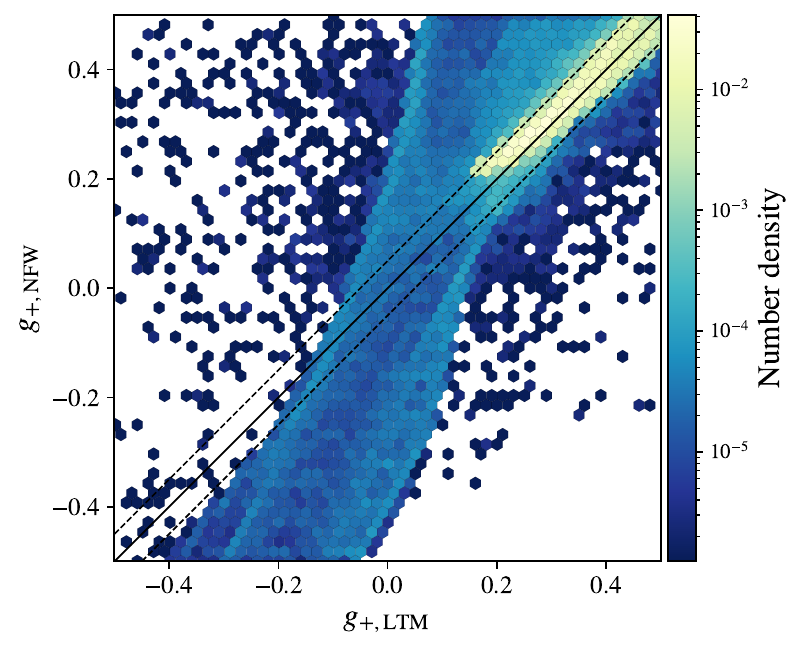}
    \caption{Tangential shear prediction from the LTM and NFW methods. The histogram shows the distribution of pixel values with $-0.5\leq g_+\leq0.5$ and within $2R_\mathrm{ein}\approx40^{\prime\prime}$ from the cluster center; here $R_\mathrm{ein}$ is the effective Einstein radius. The solid and dashed black lines denote the $g_{+,\mathrm{LTM}}=g_{+,\mathrm{NFW}}$ relation and a $\pm10\%$ deviation from the said relation, respectively. This figure shows that the two methods predict inconsistent shear for a significant fraction of pixel locations. The scatter in the shear predictions is also likely to be an underestimate as the two methods share several modeling assumptions.}
    \label{fig:shear_map_histogram}
\end{figure}

\bsp	
\label{lastpage}
\end{document}